\documentclass{pasa}%

\usepackage{graphicx}
\usepackage{txfonts}
\usepackage{hyperref}
\usepackage{times}
\usepackage{url}
\usepackage[usenames]{color}
\DeclareGraphicsExtensions{.pdf,.png,.jpg,.mps,.eps,.ps}
\usepackage{amsmath, comment}
\usepackage{multirow, float}
\usepackage{natbib}
\usepackage{caption}
\usepackage{subcaption}
\usepackage{soul}
\usepackage{lscape}
\usepackage{longtable}
\usepackage[utf8]{inputenc}

\newcommand{\angstrom}{\text{\normalfont\AA}}

\newcommand\swift{{\it Swift}}

\newcommand\astrosat{{\it AstroSat}}
\newcommand\xmm{{\it XMM-Newton}}

\newcommand\s{{\rm~s}}

\newcommand\kev{{\rm~keV}}

\newcommand\xiunit{\ifmmode {\rm~erg\s}$^{-1}$ \else ~erg~cm~s$^{-1}$\fi}
\newcommand\kms{\ifmmode {\rm~km\ s}$^{-1}$ \else ~km s$^{-1}$\fi}
\newcommand\Hunit{\ifmmode {\rm~km\ s}$^{-1}$\ {\rm Mpc}$^{-1}$
        \else ~km s$^{-1}$ Mpc$^{-1}$\fi}
\newcommand\ctssec{\ifmmode {\rm~count\ s}$^{-1}$ \else ~count s$^{-1}$\fi}
\newcommand\ergsec{\ifmmode {\rm~erg\ s}$^{-1}$ \else
        ~erg s$^{-1}$\fi}
\newcommand\funit{\ifmmode {\rm~erg\ s}$^{-1}$\;{\rm cm}$^{-2}$ \else
        ~ergs s$^{-1}$ cm$^{-2}$\fi}
\newcommand\phflux{\ifmmode {\rm~photon\ s}$^{-1}$\;{\rm cm}$^{-i2}$
        \else   ~photon s$^{-1}$ cm$^{-2}$\fi}
\newcommand\efluxA{\ifmmode {\rm~erg\ s}$^{-1}$\;{\rm cm}$^{-2}$\;{\rm
        \AA}$^{-1}$ \else ~erg s$^{-1}$ cm$^{-2}$ \AA$^{-1}$\fi}
\newcommand\efluxHz{\ifmmode {\rm~erg\ s}$^{-1}$\;{\rm cm}$^{-2}$\;{\rm
        Hz}$^{-1}$ \else ~erg s$^{-1}$ cm$^{-2}$ Hz$^{-1}$\fi}
\newcommand\cc{\ifmmode {\rm~cm}$^{-3}$ \else cm$^{-3}$\fi}
\newcommand\FWHM{\ifmmode {\rm~FWHM} \else ${\rm~FWHM}$\fi}
\newcommand\Msun{\ifmmode M_{\odot} \else $M_{\odot}$\fi}
\newcommand\Lsun{\ifmmode L_{\odot} \else $L_{\odot}$\fi}

\newcommand\hbeta{\ifmmode {\rm H}\beta \else H$\beta$\fi}
\newcommand\Kalpha{\ifmmode {\rm K}\alpha \else K$\alpha$\fi}
\newcommand\nh{\ifmmode N_{\rm H} \else N$_{\rm H}$\fi}

\title[Mrk509]{Complex optical/UV and X-ray variability in Seyfert~1 galaxy Mrk~509}

\author[Kumari et al.]{Neeraj Kumari$^{1,2}$\thanks{neerajkumari@prl.res.in, neerjakumari108@gmail.com}, Main Pal$^3$, Sachindra Naik$^1$, Arghajit Jana$^1$, Gaurava K. Jaisawal$^4$, and Pankaj Kushwaha$^5$
\affil{$^1$Astronomy and Astrophysics Division, Physical Research Laboratory, Ahmedabad-380009, India}%
\affil{$^2$Indian Institute of Technology, Gandhinagar-382355, India}
\affil{$^3$Centre for Theoretical Physics, Jamia Millia Islamia, New Delhi-110025, India}
\affil{$^4$National Space Institute, Technical University of Denmark, Elektrovej 327-328, DK-2800 Lyngby, Denmark}
\affil{$^5$Aryabhatta Research Institute of Observational Sciences, Manora Peak, Nainital-263001, India}
}%

\jid{PASA}
\doi{10.1017/pas.\the\year.xxx}
\jyear{\the\year}

\usepackage{aas_macros}
\usepackage{hyperref} 
\hypersetup{colorlinks,citecolor=blue,linkcolor=blue,urlcolor=blue}
\usepackage[all]{hypcap} 

\begin{document}

\begin{frontmatter}
\maketitle

\begin{abstract}
We performed a detailed spectral and timing analysis of a Seyfert~1 galaxy Mrk~509 using data from the {\it Neil Gehrels Swift} observatory that spanned over $\sim$13 years between 2006 and 2019. To study the variability properties from the optical/UV to X-ray emission, we used a total of 275 pointed observations in this work. The average spectrum over the entire duration exhibits a strong soft X-ray excess above the power-law continuum. The soft X-ray excess is well described by two thermal components with temperatures of kT$_{\rm bb1}\sim$120 eV and kT$_{\rm bb2}\sim$460 eV. The warm thermal component is likely due to the presence of an optically thick and warm Comptonizing plasma in the inner accretion disk. The fractional variability amplitude is found to be decreasing with increasing wavelength, i.e. from the soft X-ray to UV/optical emission. However, the hard X-ray (2-8 keV) emission shows very low variability. The strength of the correlation within the UV and the optical bands (0.95-0.99) is found to be stronger than the correlation between the UV/Optical and X-ray bands (0.40-0.53). These results clearly suggest that the emitting regions of the X-ray and UV/optical emission are likely distinct or partly interacting. Having removed the slow variations in the light curves, we find that the lag spectrum is well described by the 4/3 rule for the standard Shakura-Sunyaev accretion disk when we omit X-ray lags. All these results suggest that the real disk is complex, and the UV emission is likely reprocessed in the accretion disk to give X-ray and optical emission.
\end{abstract}

\end{frontmatter}


\section{Introduction}
\label{sec:intro}

Active galactic nuclei (AGNs) are strongly believed to be fuelled by accreting matter from the surrounding medium around the supermassive black hole (SMBH) at the center. This has been recently confirmed through the observations of a nearby radio-galaxy M87 with the Event Horizon Telescope (EHT; \citealt{akiyama2019first}). Imaging the central black hole and the accretion disk at the center of other galaxies that are farther away is impossible even with EHT. There are, however, other indirect methods that are used to derive the physical structure and size of these objects. Continuum reverberation mapping (RM; \citealt{1982ApJ...255..419B, 2014SSRv..183..253P}) technique is one such method in which the estimated time lag between the continuum emission in shorter and longer wavelength ranges has been used to derive the size of the broad-line region (BLR) and hence the black hole virial mass in several AGNs \citep{Peterson_2004, 2009ApJ...705..199B}. Understanding of the emission mechanism in AGNs has been a topic of intense research for several decades. However, the origin of emission in different wavelength ranges of the electromagnetic spectrum and correlation between them remain ambiguous to date. The general picture is that the low energy photons (seed photons) in ultraviolet (UV) and optical bands are believed to be originated from the accretion disk and the broad-line region (BLR). These photons get inverse Compton scattered in the hot electron plasma (corona), causing X-ray emission from the AGNs \citep{1991ApJ...380L..51H}. 

The origin of the UV/optical and X-ray variabilities and correlation between them have been one of the most intriguing questions in the studies of the physical size and structure of AGNs. In Seyfert galaxies, the energy spectrum is dominated by UV radiation which is believed to be originated from a multi-colour blackbody disk \citep{1999PASP..111....1K}. The X-ray/UV/optical emission from the accretion disk shows variabilities over timescales ranging from a few hours to years in the AGNs with black hole mass in $10^6-10^9~M_\odot$ range. Although the cause of the observed variabilities is not very well understood, several possible explanations are presented to interpret the results. The first and the obvious one is the fluctuations in the mass accretion rate \citep{2008MNRAS.389.1479A}. However, fluctuation in the mass accretion rate is insufficient to explain the variabilities on short time scales (i.e., hours to days) as these fluctuations drift on a viscous time scale, which is in years for AGNs. In many Seyfert galaxies, variations in X-rays have been found to lead UV/optical variations on short-time scales \citep{10.1093/mnras/stu1636}, which can not be explained by density fluctuations. Fluctuations in X-ray leading over the UV/optical band is explained by the so-called "reprocessing model". In this model, X-ray radiation from the central corona directly illuminates the accretion disk, thereby increasing its temperature. Heating of the accretion disk due to radiation from the corona makes the disk a source of enhanced radiation in UV/optical bands. In this process, the time delay between the driving radiation and reprocessed radiation is given by $\tau \propto \lambda^{\beta}$, where $\beta=4/3$ for the standard Shakura-Sunyaev accretion disk \citep{2007MNRAS.380..669C, 10.1093/mnras/stu1636, 2015ApJ...806..129E,2016ApJ...821...56F,df56ad58f9a3423b849b15173f292c20,Edelson:2017jls}. Early studies were mainly performed with radiation in UV and optical bands and were found to be consistent with the $4/3$ dependence. 

The often seen strong temporal correlation between X-ray and UV/optical emission indicates a connection between the two. Thus, simultaneous multi-wavelength observations are necessary to probe the physical processes taking place near the central engine. Before the advent of simultaneous multi-frequency observatories, e.g. \xmm{}, {\it The Neil Gehrels Swift Observatory} (hereafter \swift) and \astrosat{}, many coordinated monitoring programs have been carried out using the space-based and ground-based observatories (\citealt{Maoz_2002,2008MNRAS.389.1479A,2009MNRAS.397.1177E,2010MNRAS.403..605B}). Coordinated observations with the {\it Rossi~X-ray~Timing~Explorer (RXTE)} and many ground-based telescopes divulged the lag of optical emission with respect to the X-rays and showed a pretty good correlation between them (\citealp{2003MNRAS.343.1341S, Uttley_2003, 2008MNRAS.389.1479A, 2008ApJ...677..880M, 10.1111/j.1365-2966.2009.15110.x}). In some other cases, however, radiation in longer wavelengths was found to be leading the shorter wavelength emission \citep{2001ApJ...554L.133P, 2004MNRAS.348..783M}.

Simultaneous multi-wavelength and high cadence capabilities of the \swift{} observatory has given a thrust to the field of variability studies of AGNs. Originally designed for and focused on mainly gamma-ray bursts (GRBs), \swift{} has been utilized for the RM studies of many Seyfert~1 galaxies \citep{refId0, 10.1093/mnras/stu1636, 2015ApJ...806..129E, 10.1093/mnras/stw2486, Noda_2016, df56ad58f9a3423b849b15173f292c20, 10.1093/mnras/stw3173, 10.1093/mnras/stw878, https://doi.org/10.1002/asna.201612337, 2016ApJ...821...56F, Edelson:2017jls, Starkey_2017} and revealed the correlation between the radiation in UV/optical and X-ray bands, corroborating the reprocessing scenario. However, the observed correlation between X-ray and UV/optical emission in most cases has been found to be weak compared to that between the UV and optical emission \citep{Edelson_2019}. In several AGNs, the time lag has been found to be $\sim2-3$ times larger than the lag expected from the standard accretion disk model \citep{10.1093/mnras/stu1636}. This indicates that there are other reprocessing regions apart from the accretion disk. Excess lag in the U-band of \swift{} containing the Balmer jump compared to either X-ray or extreme UV band hints towards the contribution of BLR \citep{2001ApJ...553..695K, 10.1093/mnras/stz2330}. Earlier results from the \swift{} observations revealed that the extrapolation of the measured lag-spectrum, following the $\tau \propto \lambda^{4/3}$ relation down to X-rays, showed large deviation \citep{Dai_2009, Morgan_2010, Mosquera_2013, 2015ApJ...806..129E, 2016ApJ...821...56F, 10.1093/mnras/sty1983}. Possible explanations for this excess lag could be (i) the size of the accretion disk is larger than that predicted by standard disk \citep{10.1093/mnras/stw3173}, (ii) reprocessing of FUV photons in the inner "puffed-up" Comptonized disk region \citep{10.1093/mnras/stx946}, (iii) reprocessing of X-rays from scattering atmosphere \citep{1996ApJ...462..136N}. On filtering out the longer timescale variabilities from the light curves, the excess lags are found to decrease significantly, making the X-ray and UV/optical correlation stronger. This behaviour has been found in NGC~5548 \citep{2015ApJ...806..129E}, NGC~4593 \citep {10.1093/mnras/sty1983}, NGC~7469 \citep{10.1093/mnras/staa1055}. 


The paper is organized as follows: In Section~\ref{sec:obs-data}, the observation details and data reduction processes have been discussed. Spectral fitting, correlation analysis between original light curves, filtering of the light curves for long term variations and corresponding results are described in Section~\ref{sec:results}. Finally, the discussion on our findings has been presented in Section~\ref{sec:summary}. Throughout our analysis, we used the cosmological parameters $H_0=71\rm~km~s^{-1}~Mpc^{-1}$, $\Omega_m=0.27$ and $\Omega_{\Lambda}=0.73$.

\begin{figure*} 
\centering
\includegraphics[scale=0.8, trim={0 0.6in 0 0}, clip]{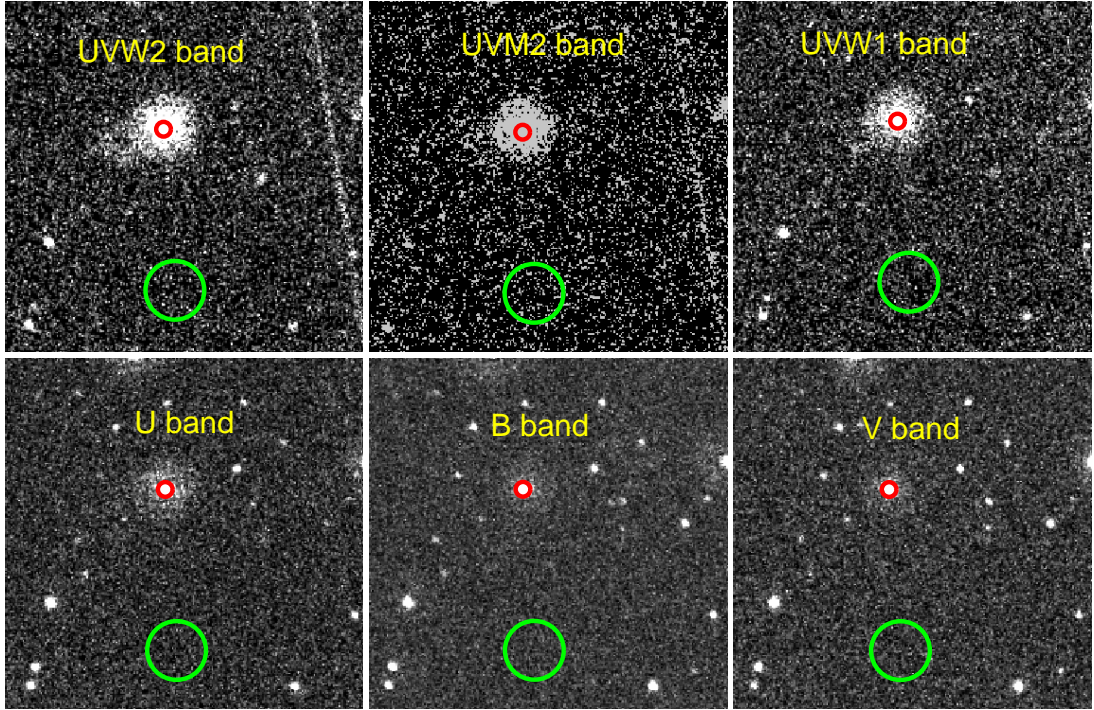} 
\caption{Images of Mrk~509 in six filters of UVOT (marked in each panel) are shown for observation ID : 00035469003. Circular regions of 5 arcsec (red, centered at source position) and 20 arcsec (green, away from the source) radii were selected for the source and background regions, respectively.}
\label{obs-net1} 
\end{figure*}

\section{Observations and data reduction}
\label{sec:obs-data}
In this work, we used archival data of Mrk~509 from \swift{} observatory that spanned over 2006 July 18 to 2019 May 8. The observations were carried out simultaneously in UV/optical and X-ray bands with the UltraViolet-Optical Telescope (UVOT; \citealt{Roming2005}) and X-ray Telescope (XRT; \citealt{2005SSRv..120..165B}), respectively. The UV/optical observations were carried out using six filters viz. UVW2 ($192.8\pm65.7$ nm), UVM2 ($224.6\pm49.8$ nm), UVW1 ($260\pm69.3$ nm), U ($346.5\pm78.5$ nm), B ($439.2\pm97.5$ nm), and V  ($546.8\pm76.9$ nm) \citep{10.1111/j.1365-2966.2007.12563.x}, whereas the X-ray observations were obtained in 0.3-10 keV range. The details of the observations used in the present work are listed in Table~\ref{obslog}. 

We extracted light curves in different energy ranges and spectra using the online XRT product generator tool\footnote[1]{http://www.swift.ac.uk/user\_objects/}. The method used in the online tool is described in \citet{2007A&A...469..379E, 2009MNRAS.397.1177E}. For generating light curves and spectra, each event file was divided into individual snapshots and further in time intervals for pile-up correction. The source extraction radius is chosen between 11.8 to 70.8 arcsec depending on the mean-count rate. The pile-up correction was performed on the time intervals in photon counting (PC) mode where mean count rate is greater than 0.6 counts~s$^{-1}$, by fitting the wings of the source PSF with the King function. For each interval, a source event list and an Ancillary Response File (ARF) were generated and then combined. While combining, each ARF was weighted according to the total counts in the source spectrum extracted from that particular time interval. The background spectrum was extracted by selecting an annular region with inner and outer radii of 142 arcsec and 260 arcsec, respectively, within the detector window. Any other source in the background region was excluded from the extraction region. The average count rate was calculated for each observation ID in PC mode of XRT for further timing and spectral analysis of the source.

Standard procedure was followed to reduce the UVOT data. We used aspect corrected sky image files of each observation ID in every filter to get the average count rate for the selected source region. A circular region of 5 arcsec radius, centered at the source, was selected as the source region to minimize contamination from the host galaxy. For background products, a circular region of 20 arcsec was selected away from the source to avoid contamination from the host galaxy as well as from any other source in the field. The regions selected for the source and background are shown in red and green circles in Figure~\ref{obs_net1}, respectively. The exposures of the UVOT filters are generally in the range of 10~s to 2~ks. To get the deepest image, {\tt UVOTISUM}\footnote[2]{https://www.swift.ac.uk/analysis/uvot/image.php} task was used to co-add multiple exposures, if available, in each observation ID. The {\tt UVOTSOURCE}\footnote[3]{https://www.swift.ac.uk/analysis/uvot/mag.php} task was used to get the background-corrected source count rate/flux densities in different filters. We also checked for small scale sensitivity (SSS) regions where throughput of the detector is comparatively low by running {\tt UVOTSOURCE} task with additional command {\tt lssfile=sssfile5.fits.} Apart from this, there are a number of data points where the count rate/flux fall rapidly and show large variability compared to the local mean. This may be due to the bad tracking of the telescope, as described in \citet{10.1093/mnras/stu1636}. We discarded those data points from further analysis. We also corrected the count rates to account for the loss of sensitivity of UV detectors with time by using filter correction files\footnote[4]{https://www.swift.ac.uk/analysis/uvot/index.php}.

\begin{table}
   \noindent \begin{centering}
     \caption{Log of observations of Mrk~509 with the \swift{} observatory} \label{obs-log}
     \begin{tabular}{lr}
       \hline 
 Observation ID           & 00035469001--00095002006\\
 Duration of observation  & 2006 July 18 - 2019 May 8\\
 MJD                      & 53812.05 - 58611.70\\
 No. of IDs for XRT used  & 275\\
 No. of IDs for UVOT used & 271\\
        \hline
     \end{tabular}
     \par\end{centering}
 \end{table}

--------------------------------------------------------------------------------------------------
\section{Data Analysis \& Results}
\label{sec:results}

\subsection{ Spectral analysis}
\label{subsec:spectrum}

 \begin{figure}
 	\centering
 	\includegraphics[scale=0.35, angle=-90]{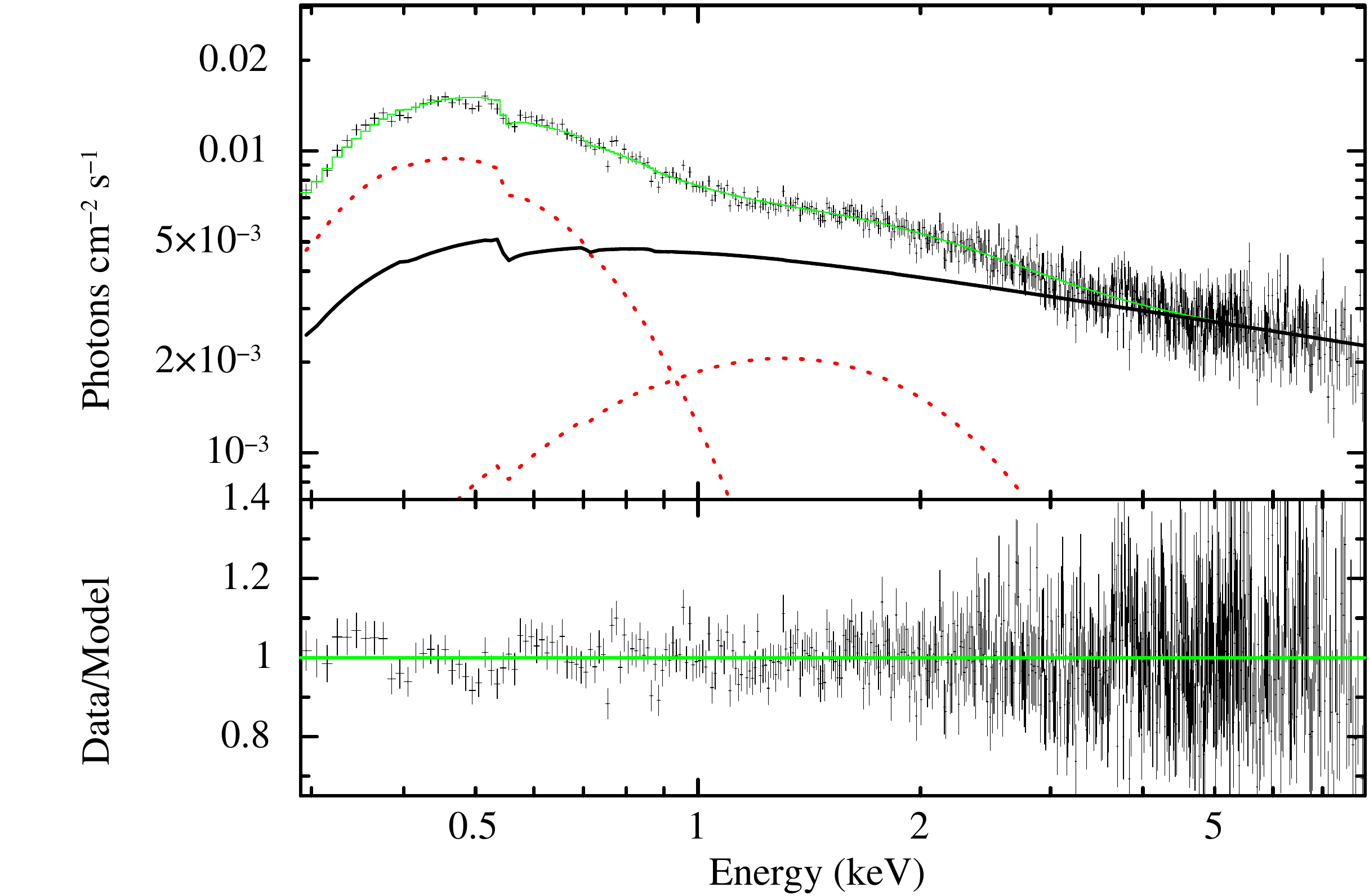}
 	\caption{Time-averaged \swift/XRT spectrum of Mrk~509 and best-fit model (green line) are shown along with individual spectral components (red dotted lines for two blackbody components and solid black line for the power-law continuum model) in the top panel. Corresponding residuals are shown in the bottom panel.}
 	\label{bestfit}
 \end{figure}

The AGNs spectra, especially of Seyferts, in 0.3-10 keV range (the range of operation of \swift/XRT) generally consist of soft X-ray excess and a power-law continuum along with the iron emission lines in $\sim$6-7 keV range. In order to investigate the origin of the soft X-ray excess and power-law continuum, we carried out spectral analysis using data from all the \swift/XRT observations of Mrk~509. We extracted the time-averaged spectrum of all the observation IDs using the online XRT product generator tool\footnote[1]{http://www.swift.ac.uk/user\_objects/} \citep{2009MNRAS.397.1177E}. We fitted the spectrum in {\tt XSPEC} and used C-statistics for minimization to obtain the best-fit parameters. The errors on each parameter are quoted at 90\% confidence level. Using the source and background spectra and appropriate response matrices, we attempted to fit the data in 2-8 keV range with a redshifted {\tt power-law} model along with the multiplicative component {\tt tbabs} to incorporate the modification due to the Galactic absorption. In our fitting, the Galactic column density was fixed at $3.95\times10^{20}$ cm$^{-2}$ \citep{2016A&A...594A.116H}. The fit statistic was found to be $C/dof=580.8/597$ for a best-fit power-law photon index of $1.68\pm0.02$. We extrapolated the fitted model down to 0.3 keV. This showed strong positive residuals in the soft X-ray range (below 1 keV). This confirms the presence of soft X-ray excess in the spectrum. The soft X-ray excess above the power-law continuum was fitted with a simple redshifted blackbody ({\tt zbbody}) model. The positive residuals were still present up to $\sim$3 keV. We added another redshifted blackbody component to the above model to describe these residuals. The fit-statistic improved by $\Delta C=-91$ with two additional parameters. The best-fit model {\tt tbabs$\times$(zbbody+zbbody+zpowerlaw)} resulted in $C/dof=730.1/764$. The best-fit value of the power-law photon index was found to be $\Gamma=1.39\pm0.04$, which is consistent with that reported by \citet{2011A&A...534A..39M}. The best-fit model, data and residuals are shown in Figure~\ref{bestfit}. The best-fitted parameters are given in Table~\ref{fit-par}.

 \begin{table}
 	\centering
 	\caption{Best-fit parameters obtained from spectral fitting of \swift/XRT data.}
 	\label{fit-par}
 	\begin{tabular}{ll}
 		\hline 
 		Spectral parameter & Value    \\
 		\hline
 		Absorption column density N$_{\rm H}$ [10$^{20}$ cm$^{-2}$] & 3.95 (f) \\
 		Blackbody Temperature kT$_{\rm BB1}$ [keV] & 0.121$\pm0.002$\\
 		Blackbody Normalization kT$_{\rm norm1}$ [10$^{-5}$]  & 17.4$\pm0.8$\\
 		
 		Blackbody Temperature kT$_{\rm BB2}$ [keV] & 0.46$\pm0.01$\\
 		Blackbody Normalization kT$_{\rm norm2}$ [10$^{-5}$]  & 9.0$\pm1.0$\\
 		Power-law Photon Index $\Gamma$ & 1.39$\pm0.04$ \\
 		Power-law Normalization $\Gamma_{\rm norm}$ [10$^{-3}$] & 5.3$\pm0.4$ \\
 		$C$/dof & 730.1/764 \\
 	\end{tabular}
 	\vspace*{3mm}
 	\leftline{Best fit model : {\tt tbabs $\times$ [(zbbody+zbbody+zpowerlw)]}. } 
 \end{table}

Following the time-averaged spectroscopy, we attempted to fit the spectra from individual observation IDs with exposures of $\sim$1 ks. Each spectrum was grouped at a minimum of 1 count per bin to use C-statistics in {\tt XSPEC} spectral fitting. We fitted each spectrum with a simple phenomenological model {\tt tbabs$\times$(bbody+zpowerlw)} as the second thermal component was not required in fitting. While fitting the spectrum of individual IDs with the above model, we estimated source flux in 0.3-8 keV, 0.3-2 keV, and 2-8 keV ranges for further analysis. The total flux in 0.3-8 keV range, best-fit spectral parameters and the reduced $C$ for all observation IDs are shown in Figure~\ref{KT_index}. The variation/non-variation of flux and spectral parameters over the duration of the observations can be clearly seen in the figure. We attempted to search for the presence of any correlation between different parameters obtained from spectral fitting. The correlations between the power-law flux (in 2-8 keV range), blackbody flux (BB flux in 0.3-2 keV range), power-law photon index and blackbody temperature are shown in  Figure~\ref{par_corr}. From this analysis, we found a moderate correlation between the blackbody flux and power-law flux with Pearson's coefficient $\rho\sim0.56$, between the blackbody temperature and blackbody flux with $\rho\sim0.31$, and between the blackbody temperature and photon index with $\rho\sim-0.38$. Considering the narrow energy range of XRT and short exposure of observations, there is a possibility of presence of degeneracy between parameters. Therefore, we investigated the marginal posterior distributions of all the fitted parameters to determine the degree to which the parameters are correlated due to the degeneracies within the spectral fitting. In order to search through the parameter space, the data from the observation with maximum exposure, i.e. $\sim$7 ks (Obs. ID 00035469003) were used for fitting by applying the Markov Chain Monte-Carlo (MCMC) sampling procedure. We used the affine-invariant sampler developed by \citet{2010CAMCS...5...65G} and implemented in {\tt XSPEC} as {\tt CHAIN} task. For {\tt tbabs$\times$(zbbody+zpowerlw)} model, we considered 20 walkers with a total chain length of 100000 and burning initial steps of 20000. In Figure~\ref{mcmc}, we showed one and two-dimensional (1D and 2D) marginal posterior distributions for the spectral parameters. The median values of the parameters with 90\% credible interval are shown above each 1D histogram. The Pearson's correlation coefficients for these distributions are $-0.26$ for kT$_{\rm BB}$ \& kT$_{\rm norm}$, $-0.49$ for $\Gamma$ \& kT$_{\rm BB}$, $-0.47$ for $\Gamma$ \& kT$_{\rm norm}$, $-0.59$ for $\Gamma_{\rm norm}$ \& kT$_{\rm BB}$, $-0.44$ for $\Gamma_{\rm norm}$ \& kT$_{\rm norm}$, and $0.85$ for $\Gamma_{\rm norm}$ \& $\Gamma$ with $p$-value <$10^{-5}$. It is evident from the figure that the correlation between the blackbody temperature and the photon index is not physical, instead spurious. On a careful investigation of correlation between spectral parameters from spectral fitting of all the observations and the MCMC sampling procedure, it is confirmed that there exists a certain degree of degeneracy between power-law photon index and blackbody temperature. However, the narrow energy range of {\it Swift}/XRT and short exposure time ($\sim$ 1 ks) of individual observations make it extremely difficult to remove this degeneracy through spectral fitting. Considering this, we proceeded with time-series analysis using data from all the {\swift}/XRT observations.
\begin{figure*} 
 	\includegraphics[width=19cm, height= 10cm, trim={0 0 0 0}]{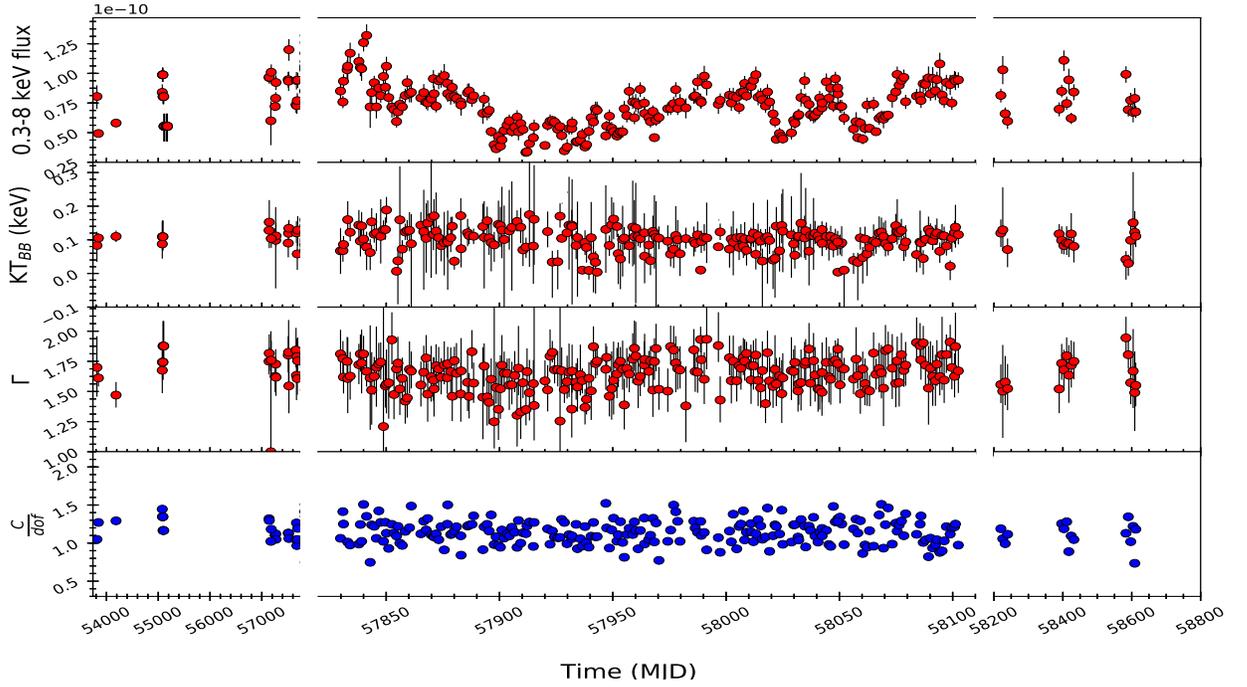}
 	\caption{Variation of the source flux in 0.3--8 keV range, blackbody temperature (kT) and power-law photon index ($\Gamma$) with time (MJD) are shown in top three panels. The reduced $C$-stat obtained from the spectral fitting of each observation ID is plotted in the bottom panel.}
 	\label{KT-index} 
 \end{figure*}

 \begin{figure}
 	\centering
	\includegraphics[scale=0.55, trim={0.5in 0 0 2.0in}]{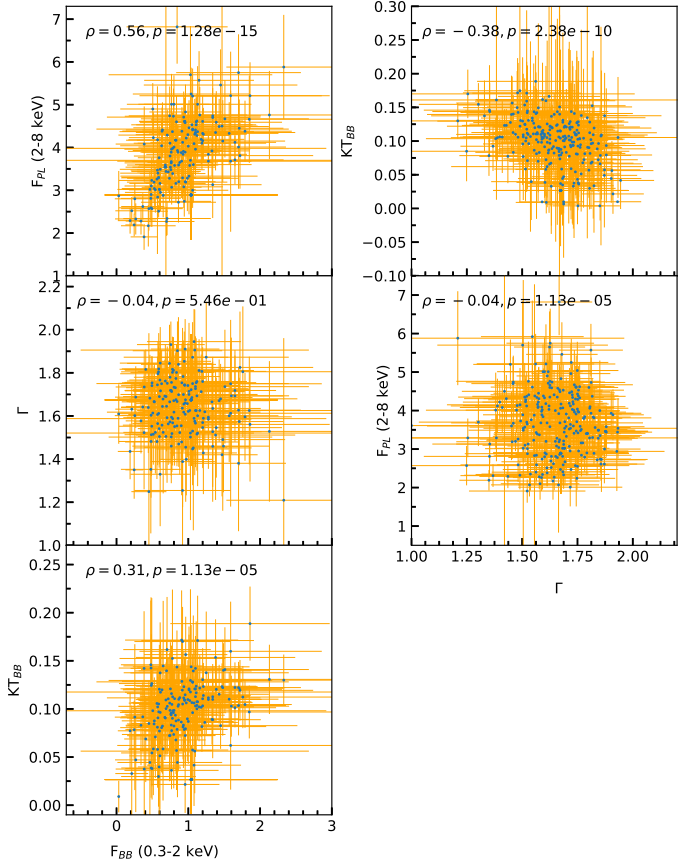}
 	\caption{Pearson's correlation coefficients ($\rho$, p-value) between different spectral parameters extracted from fitting of individual observations using phenomenological model {\tt tbabs$\times$[bbody+zpowerlw]}. The 0.3-2 keV blackbody flux (F$_{\rm BB}$; in units of 10$^{-11}$ erg s$^{-1}$ cm$^{-2}$) has been plotted against 2-8 keV power-law flux (F$_{\rm PL}$; in units of 10$^{-11}$ erg s$^{-1}$ cm$^{-2}$), photon-index ($\Gamma$) and blackbody temperature KT$_{\rm BB}$ (in unit of keV) from top to bottom, respectively, in the left panels of the figure. In the right panels, the plots for $\Gamma$ vs KT$_{\rm BB}$ and $\Gamma$ vs F$_{\rm PL}$ have been shown.}
 	\label{par-corr}
 \end{figure}

 \begin{figure*} 
  \centering
  \hspace*{-0.2in}
  \includegraphics[width=20cm, height=20cm, angle=0, trim={1.5cm 0 0 1.8cm}, clip]{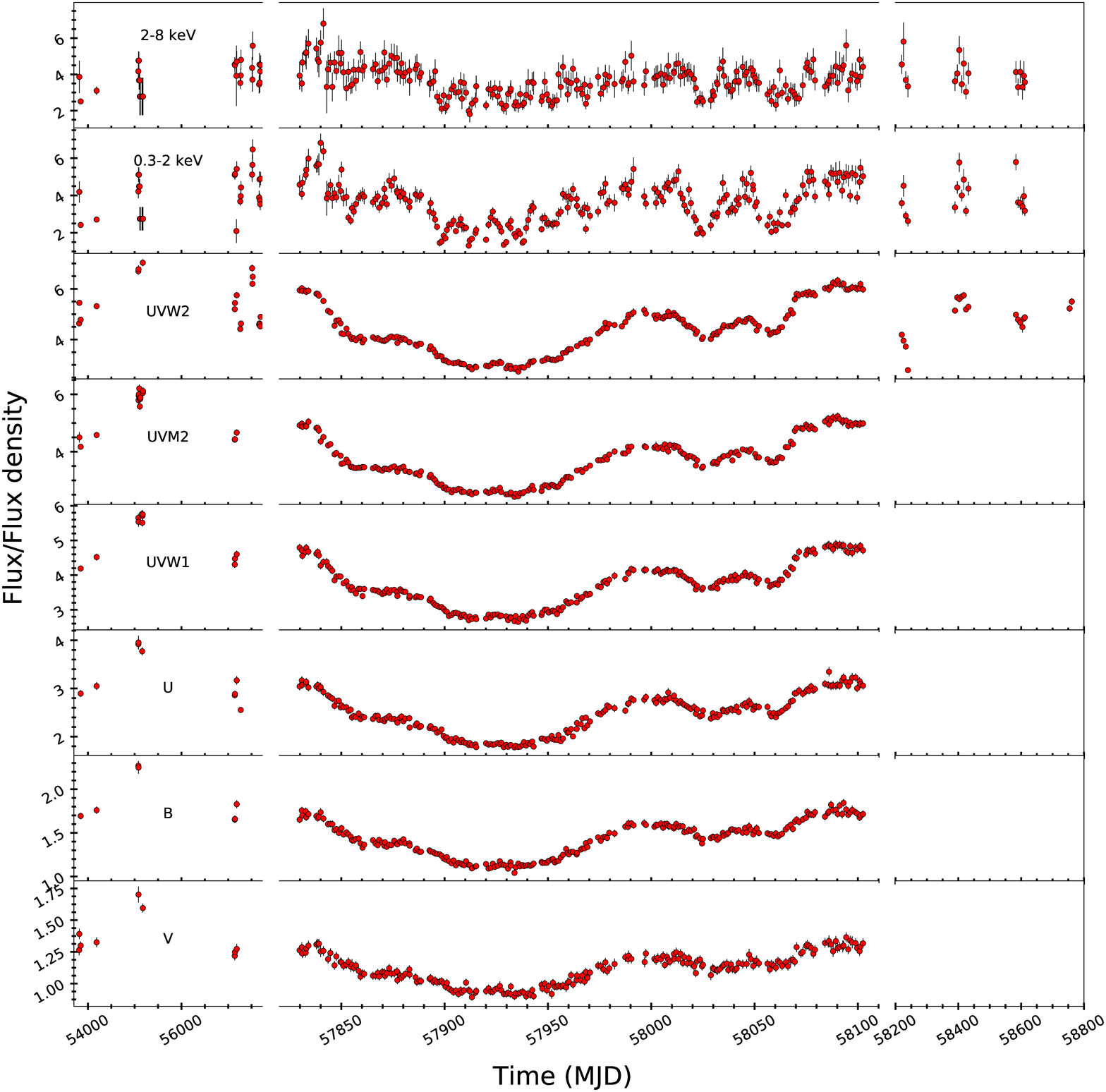}
  \caption{Simultaneous UV/optical and X-ray light curves of Mrk~509 from \swift{} observations during 2006 June 18 to 2019 May 8. The UV/optical (UVW2, UVM2, UVW1, U, B, V) flux densities (in 10$^{-14}$ erg s$^{-1}$ cm$^{-2}$ \AA$^{-1}$) of the source are plotted with the hard X-ray (2-8 keV) and soft X-ray (0.3-2 keV) flux densities (10$^{-11}$ erg s$^{-1}$ cm$^{-2}$). The X-ray flux densities were estimated from spectral fitting of data from the individual observation IDs. Two breaks in the light curves are due to the large observation gaps.}
  \label{lightcur} 
  \end{figure*}

 \begin{figure*}
      \centering
      \includegraphics[width=15cm]{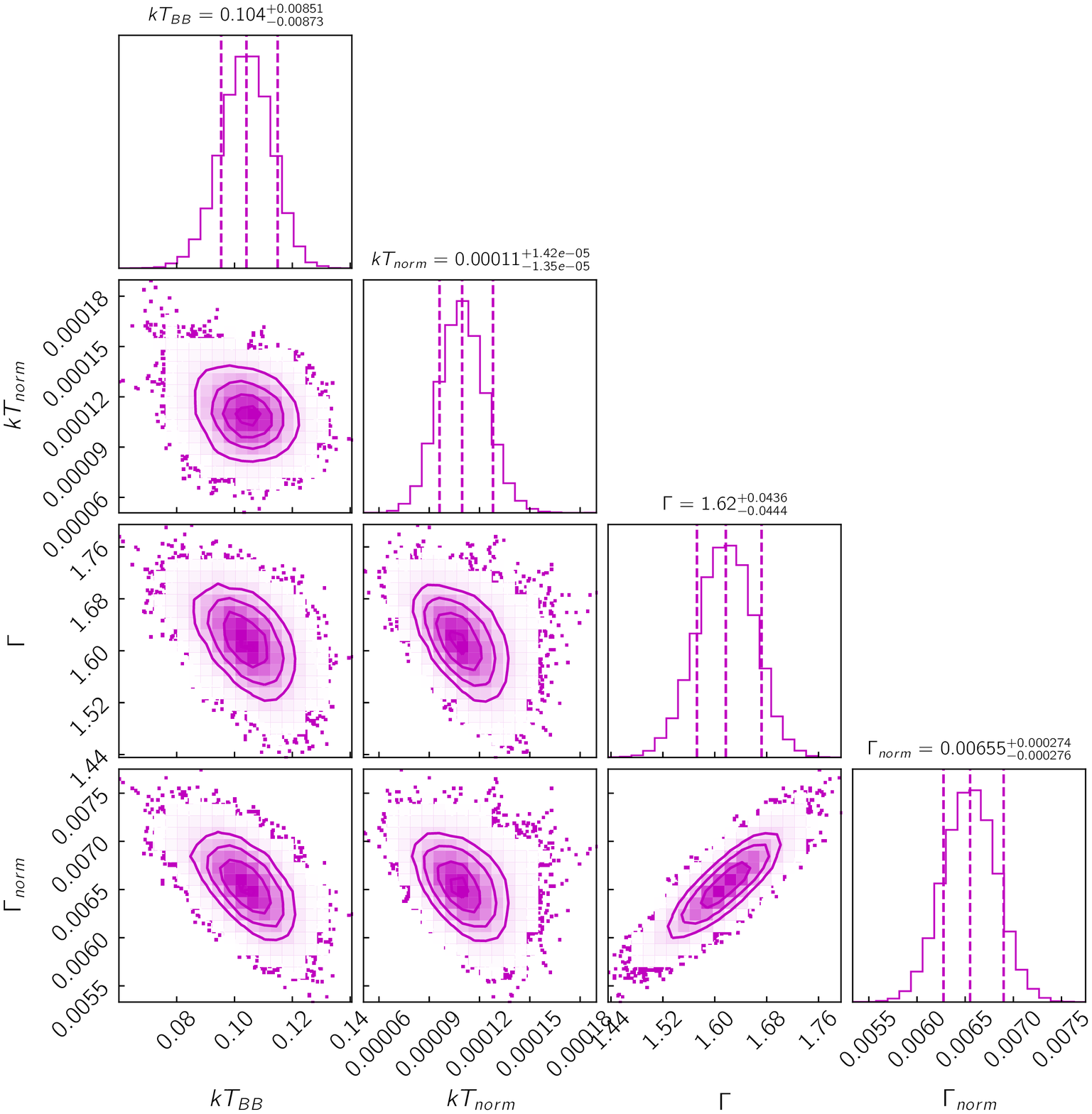}
      \caption{1D and 2D marginal posterior distributions for power-law photon index ($\Gamma$) and its normalization ($\Gamma_{\rm norm}$), blackbody temperature ($kT_{\rm BB}$, in keV) and its normalization ($kT_{\rm norm}$) for Obs. ID 00035469003 fitted with {\tt tbabs$\times$[bbody+zpowerlw]} model. Vertical lines in 1D distributions show 16\%, 50\% and 90\% quantiles. {\tt CORNER.PY} \citep{corner} was used to plot these distributions.}
      \label{mcmc}
  \end{figure*}

\subsection{Timing analysis}
\label{subsec:timing}

\subsubsection{Variability amplitudes \& nature of variability} 
\label{subsubsec: varaibility}
As described in Section~\ref{sec:obs-data}, the source flux densities for all six UVOT filters were estimated from all observation IDs. Soft X-ray (0.3-2 keV range) and hard X-ray (2-8 keV) fluxes were estimated from spectral analysis of \swift/XRT data of each observation ID (see Section~\ref{subsec:spectrum}). Using these estimated flux densities, simultaneous X-ray and UVOT light curves of Mrk~509 for the duration of the \swift{} monitoring campaign, i.e. from 2006 June 18 to 2019 May 8, are generated and shown in Figure~\ref{lightcur}. It can be clearly seen that the light curves in different bands show variabilities in short-term as well as long-term time scales. To quantify the observed variabilities in X-ray and UV/optical light curves, we calculated fractional variability $F_{\rm var}$ and its uncertainty using the relations,

  $$F_{var}=\sqrt{\frac{S^2-\overline{\sigma^2_{err}}}{\Bar{x}^2}}$$ 

 and uncertainty in $F_{\rm var}$,

  $$\sqrt{\Bigg( \sqrt{\frac{1}{2N}} \frac{\overline{\sigma^2_{\rm err}}}{\Bar{x}^2 F_{var}} \Bigg)^2 + \Bigg( \sqrt{\frac{\overline{\sigma^2_{\rm err}}}{N}} \frac{1}{\Bar{x}}\Bigg)^2 }$$

 where $\Bar{x}$, $S$, $\overline{\sigma}_{\rm err}$ and $N$ are the mean, total variance, mean error and number of data points, respectively \citep{2003MNRAS.345.1271V}, for X-ray and UV/optical bands. Using the above expressions, the fractional variabilities in hard X-ray (2-8 keV), soft X-ray (0.3-2 keV), UVW2, UVM2, UVW1, U, B, and V bands are derived to be $0.152\pm0.010$, $0.281\pm0.006$, $0.223\pm0.001$, $0.234\pm0.002$, $0.184\pm0.002$, $0.174\pm0.002$, $0.147\pm0.002$ and $0.114\pm0.002$, respectively. Within the UV/optical bands, the variability amplitudes are found to decrease with the increase in wavelength. This trend has been reported in many previous studies \citep{2016ApJ...821...56F, Edelson_2019} as emission at longer wavelengths is expected to originate from regions in accretion disk at larger radii and thus, prone to dilute with other effects like emission from BLR and narrow-line region (NLR) or host galaxy. From the spectral analysis, we found a complex soft X-ray excess described by two thermal components. Thus, high variability in the soft X-ray band could be due to a mixture of different spectral components i.e., thermal plasma in the inner disk, X-ray reprocessing close to the inner edge of the disk. Since the signatures of warm absorbers have already been reported in previous studies \citep{refId0, 10.1051/0004-6361/201117304}, and we see some residuals below 1 keV 

\begin{figure*}
      \centering
      \includegraphics[scale=0.38, trim={1.5cm 0 0 1.4cm}]{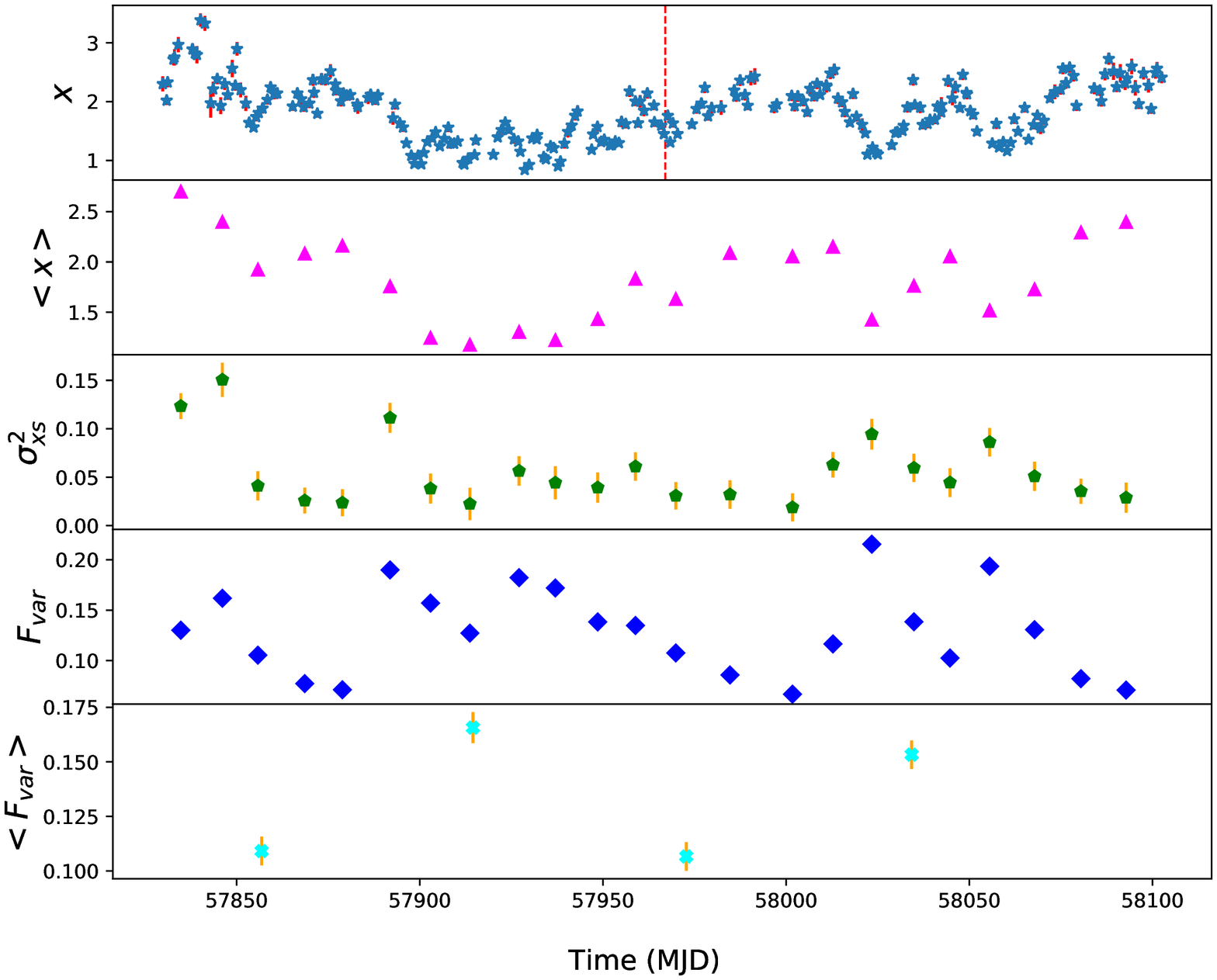}
      \includegraphics[scale=0.378, trim={1.5cm 0 0 1.4cm}]{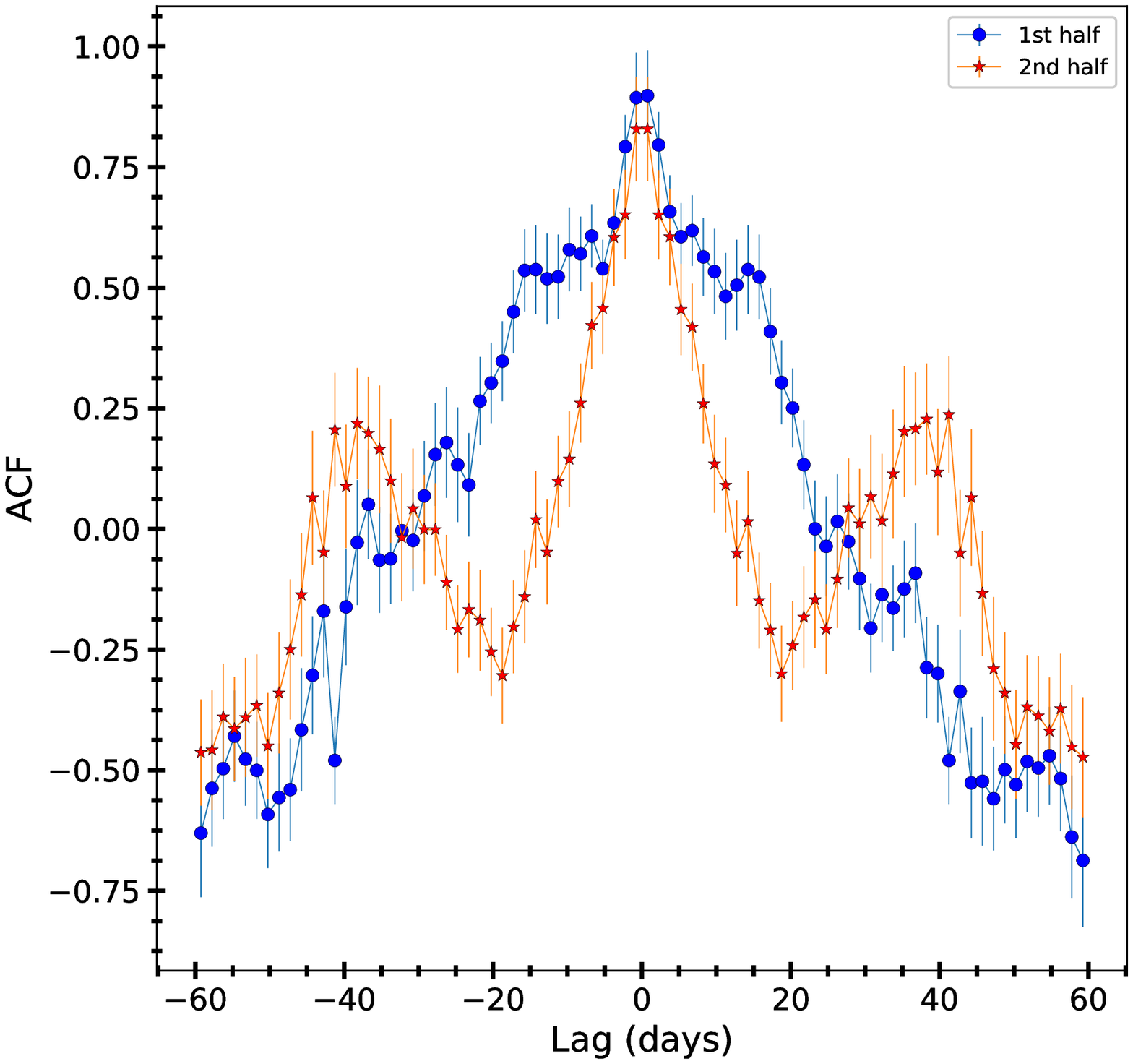}
      \caption{Left panel: Light curve of Mrk~509 in 0.3--10 keV range is shown in the top panel. The mean count rate, excess variance and fractional rms amplitude measured from the segment of 10 points are shown in the 2nd, 3rd and 4th panel from top. In the bottom panel, averaged fractional rms amplitude is shown by binning 5 amplitudes. Right panel : Auto-correlations calculated for two segments (from 57829 to 57967 MJD and 57968 to 58102 MJD, as marked with a dotted line in the top left panel) of 0.3-10 keV light curves are shown.}
      \label{stationarity}
  \end{figure*}

Based on the probability spectral density (PSD) studies, the nature of variability in AGNs is considered to be red-noise variability \citep{1993MNRAS.261..612P}. This is very similar to the one that is observed in Galactic black holes, where the PSD is described by a broken power-law at a specific frequency depending on the spectral state. Switching between different states indicates that the variability process is non-stationary. Statistically, in non-stationary process, the moments of the probability distribution function (i.e. mean and variance) vary with time. In the left panel of 
we plotted the change of mean $<x>$, excess variance $\sigma^2_{\rm xs}$ and fractional rms amplitude $F_{\rm var}$ (computed by taking 10 data points) in 0.3-10 keV light curve with time. All these quantities are found to change with time. In the bottom panel, the averaged rms has been shown (binned over 5 data points of individual rms), which also changes with time, indicating the non-stationarity of the variability process. We divided the entire light curve into two segments covering the time from 57829 to 57967 MJD and 57968 to 58102 MJD, respectively (marked with a dotted line in the top left panel of Figure~\ref{stationarity})
and calculated the auto-correlation function (ACF) of each segment \citep{2004HEAD....8.0402G}. The rate at which the ACF decays for each segment is different. This suggests that there is an intrinsic difference in the temporal properties in both segments. This indicates non-stationary variability, though we can not claim it firmly due to the quality of the data used. We also did not find any signature of apparent state change over the duration of the observations 

\subsubsection{Count--Count Correlation with Positive Offset (C3PO)}
\label{subsubsec:c3po}
 We defined $0.3-2$\kev~and $2-8$\kev~energy ranges as soft and hard X-ray bands. The soft X-ray band have several complex features such as soft X-ray excess, emission lines, blurred reflection and absorption features, while the hard band is mainly dominated by primary continuum resulted from multiple inverse Compton scattering of seed photons with electron plasma in the corona. We plotted light curves in soft X-ray band, hard X-ray band and all the filters of optical and UV bands in Figure~\ref{lightcur}. From visual inspection, all the light curves seem to be correlated. The variations in UV/optical bands are smoother than the X-rays. 

We attempted to quantify the correlations and variabilities seen in UV/optical and X-ray light curves using C3PO method. The C3PO method was first used by \citet{2001MNRAS.321..759C} in the black hole binary system Cygnus~X-1 to find the varying and stable components in high/soft state. After that, this technique has been used in many AGNs \citep{10.1046/j.1365-8711.2003.06742.x, 10.1093/pasj/63.sp3.S925, Noda_2013, Noda_2016, 10.1093/mnras/stx3103}. We fitted a linear function $y=mx+c$ between the reference band as abscissa and secondary bands as ordinate, where $m$ and $c$ are the slope and offset, respectively. In the left and right panels of Figure~\ref{corr_all}, the hard X-ray (2-8 keV) and UVW2 bands are taken as reference bands, respectively, and all other bands are considered as secondary. For uniformity, the plots have been shown for flux densities though there was no difference in the fitted values when plotted in count rates. We also used Pearson's correlation coefficient~`$\rho$' to quantify the strength of inter-band correlation and determined the significance of the strength of the correlation. All the values of slopes, offset, Pearson's coefficient are given in Tables~\ref{c3po1} and~\ref{c3po2}. From the linear fitting, we found a reasonable correlation between hard X-rays and all UV/optical bands ($\rho~\sim~0.40-0.53$). However, the correlation between the UVW2 and other UV-optical bands is found to be stronger ($\rho~\sim~0.95-0.99$). When fitted relative to UVW2 band, a positive offset (except for soft X-rays) has been found for all other UV/optical bands, including hard X-ray. This indicates that the less variable component is possibly coming from the BLR and the host galaxy. Similar results have been drawn from the flux-flux analysis for other systems, e.g. Fairall~9 \citep{10.1093/mnras/staa2365}.

 \begin{table*}
 	\begin{minipage}{.5\linewidth}
 		\caption{Slope, offset and Pearson's coefficient for hard X-ray vs soft bands (soft X-ray \& UV/optical bands).}
 		\label{c3po1}
 		\centering
 	\begin{tabular}{p{1.5cm} |p{2.5cm} |p{1.2cm} |p{2cm}} 
 		\hline
 		Band & slope(m) \& offset(c) & $\chi^2_\nu$, dof & Pearson's coefficient$\rho, p$\\ [0.5ex] 
 		\hline\hline
 		\multirow{2}{2cm}{Soft X-ray} & $m=1.50\pm0.05$ & \multirow{2}{1.5cm}{0.61, 289} & \multirow{2}{3cm}{0.82, $1.80\times10^{-71}$}\\ & $c=(-1.79\pm0.17)\times10^{-11}$ & & \\
 		\hline
 		\multirow{2}{2cm}{UVW2} & $m=2.17\pm0.21$ & \multirow{2}{1.5cm}{2.03, 251} & \multirow{2}{3cm}{0.52, $1.15\times10^{-18}$}\\ & $c=(-2.97\pm0.70)\times10^{-11}$ & &  \\
 		\hline
 		\multirow{2}{2cm}{UVM2} & $m=1.97\pm0.21$ & \multirow{2}{1.5cm}{2.14, 246} & \multirow{2}{3cm}{0.40, $4.95\times10^{-11}$} \\ & $c=(-2.90\pm0.70)\times10^{-11}$ & &\\
 		\hline
 		\multirow{2}{2cm}{UVW1} & $m=1.51\pm0.16$ & \multirow{2}{1.5cm}{2.13, 240} & \multirow{2}{3cm}{0.44, $6.98\times10^{-13}$}\\ & $c=(-1.37\pm0.53)\times10^{-11}$ & &\\
 		\hline
 		\multirow{2}{2cm}{U} & $m=0.92\pm0.09$ & \multirow{2}{1.5cm}{2.13, 239} & \multirow{2}{3cm}{0.52, $4.64\times10^{-18}$}\\ & $c=(-0.62\pm0.30)\times10^{-11}$ & &\\
 		\hline
 		\multirow{2}{2cm}{B} & $m=0.47\pm0.05$ & \multirow{2}{1.5cm}{2.18, 237} & \multirow{2}{3cm}{0.53, $8.64\times10^{-19}$} \\ & $c=(-0.14\pm0.16)\times10^{-11}$ & &\\
 		\hline
 		\multirow{2}{2cm}{V} & $m=0.25\pm0.023$ & \multirow{2}{1.5cm}{2.16, 234} & \multirow{2}{3cm}{0.48, $3.67\times10^{-15}$}\\ &$c=(0.27\pm0.08)\times10^{-14}$ & &\\ [1ex] 
 		\hline
 	\end{tabular}\end{minipage}%
 \hspace{0.5cm}
  \begin{minipage}{.5\linewidth}
  \centering
  \caption{Slope, offset and Pearson's coefficient for UVW2 vs other UV/optical bands.}
  \label{c3po2}
 \begin{tabular}{p{1.5cm} |p{2.5cm} |p{1.2cm} |p{2cm}} 
  	\hline
  	Band & slope(m) \& offset(c) & $\chi^2_\nu$, dof & Pearson's coefficient$\rho, p$\\ [0.5ex] 
  	\hline\hline
   	\multirow{2}{2cm}{Hard X-ray} & $m=0.46\pm0.04$ & \multirow{2}{1.5cm}{2.07, 238} & \multirow{2}{3cm}{0.51, $2.41\times10^{-17}$}\\ & $c=(1.34\pm0.19)times10^{-11}$ & & \\
  	\hline
  	\multirow{2}{2cm}{Soft X-ray} & $m=0.96\pm0.05$ & \multirow{2}{1.5cm}{6.04, 238} & \multirow{2}{3cm}{0.75, $5.76\times10^{-44}$}\\ & $c=(-0.85\pm0.20)\times10^{-14}$ & & \\
  	\hline
  	\multirow{2}{2cm}{UVM2} & $m=0.79\pm0.004$ & \multirow{2}{1.5cm}{0.26, 224} & \multirow{2}{3cm}{0.99, $6.37\times10^{-241}$}\\ & $c=(0.25\pm0.02)\times10^{-14})$ & & \\
  	\hline
  	\multirow{2}{2cm}{UVW1} & $m=0.66\pm0.005$ & \multirow{2}{1.5cm}{0.38, 224} & \multirow{2}{3cm}{0.99, $4.82\times10^{-212}$}\\ & $c=(0.85\pm0.02)\times10^{-14}$ & &  \\
  	\hline
  	\multirow{2}{2cm}{U} & $m=0.43\pm0.006$ & \multirow{2}{1.5cm}{1.10, 225} & \multirow{2}{3cm}{0.98, $2.74\times10^{-150}$} \\ & $c=(0.59\pm.02)\times10^{-14}$ & &\\
  	\hline
  	\multirow{2}{2cm}{B} & $m=0.21\pm0.003$ & \multirow{2}{1.5cm}{1.35, 225} & \multirow{2}{3cm}{0.97, $8.27\times10^{-133}$}\\ & $c=(0.50\pm0.01)\times10^{-14}$ & &\\
  	\hline
  	\multirow{2}{2cm}{V} & $m=0.12\pm0.002$ & \multirow{2}{1.5cm}{0.97, 225} & \multirow{2}{3cm}{0.95, $1.01\times10^{-113}$}\\ & $c=(0.60\pm0.01)\times10^{-14}$ & &\\[1ex] 
  	\hline
  \end{tabular}
  \end{minipage} 
  \end{table*} 

 \begin{figure} 
 \centering
 \includegraphics[width=9cm]{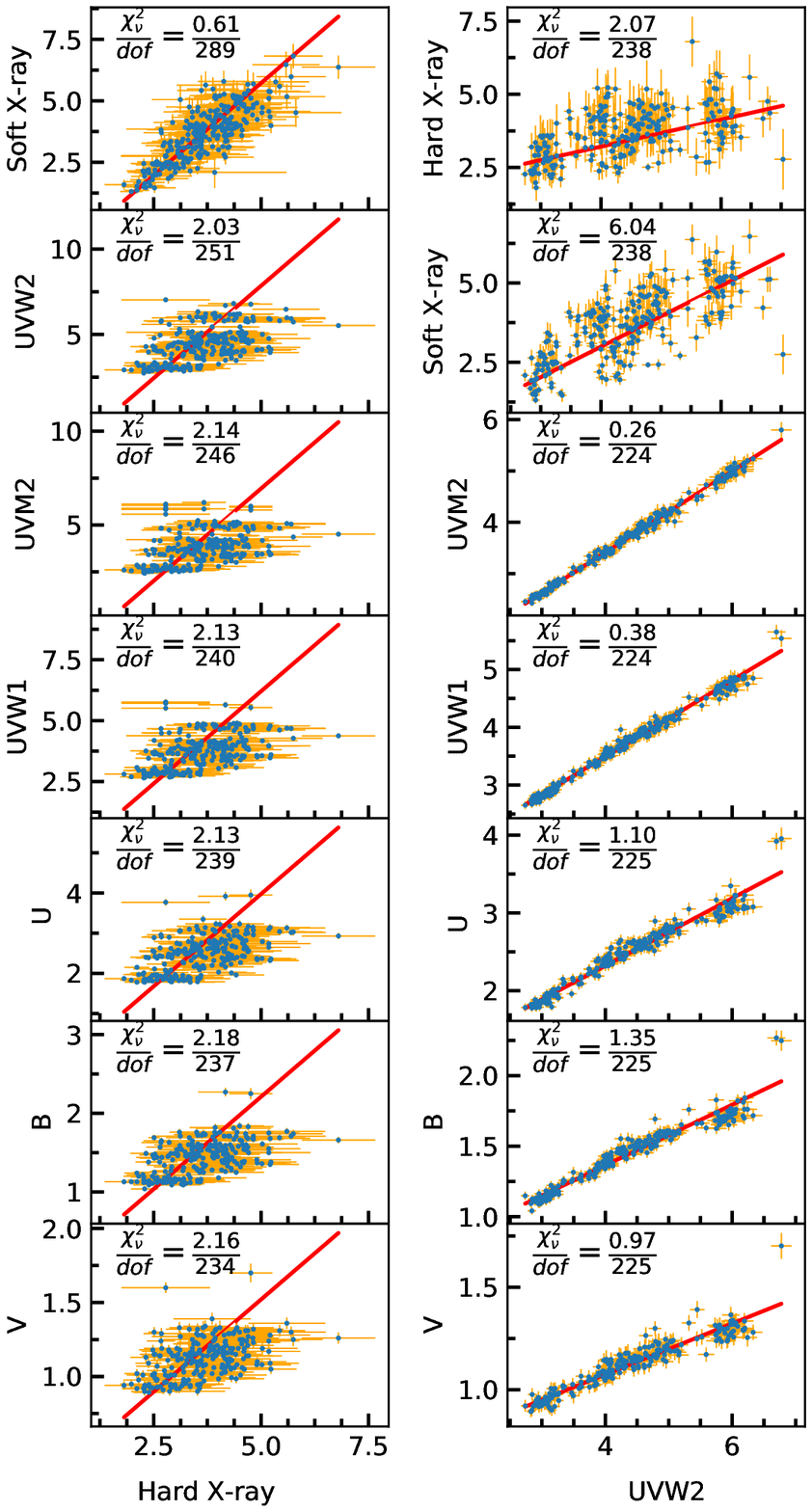}
 \caption{C3PO plots are shown for hard X-ray band (left panels) and UVW2 band (right panels) as reference bands. The best-fitted linear equations are presented as red solid lines. The units of the fluxes are 10$^{-11}$ erg s$^{-1}$ cm$^{-2}$ for X-ray bands and 10$^{-14}$ erg s$^{-1}$ cm$^{-2}$ $\AA^{-1}$ for UV/optical bands.}
 \label{corr-all} 
 \end{figure}

\subsubsection{Filtering of slow and linear variations}
\label{subsubsec:filtering}
To understand the short term variabilities in different bands due to reprocessing, i.e. X-ray reprocessing or Comptonizaton, we focused on the fast variations in the light curves. It is well known that the long term variations can distort the short term time lags while estimating CCFs \citep{1999PASP..111.1347W}. So we attempted to remove the slow variations, i.e long term variations due to the changes in the accretion rate, from the light curves. As mentioned above, Figure~\ref{lightcur} shows short term variations as well as long term variations in the curves in all bands. Therefore, the long term variations should be filtered out from the light curves to determine short time information. Such filtering has been carried out in recent studies in a few AGNs, e.g. NGC~5548 \citep{10.1093/mnras/stu1636}, NGC~4593 \citep{10.1093/mnras/sty1983}, NGC~7469 \citep{10.1093/mnras/staa1055}, to eliminate the effect of long term variations. We used the locally weighted scatter smoothing (LOWESS) method for the filtering process, which is based on non-parametric and non-linear least square regression method \citep{doi:10.1080/01621459.1988.10478639}. In this method, a regression surface is estimated by fitting a low-degree polynomial locally to a subset of the data. The polynomial is fitted with weighted least-squares, giving more weight to the near data points and least weight to the farthest ones. The tricube function, $w(s)=(1-{\lvert s \rvert}^3)^3$ is used to calculate the weight, where $s$ is the distance of a given data point from the neighbouring point scaled to lie in the range of $0$ to $1$. Here, $s(t)=(t-t_i/\Delta t)$ is the time difference between time $t$ and $t_i$ at the data point $i$ in terms of filtering time $\Delta t$. For this work, we used different filtering time to calculate the time lags as well as to investigate the effect of filtering of the long term variations. Since the filtering process has a limitation that it can only be performed on densely populated sample, we, therefore, applied the LOWESS filtering to the data in the range of MJD 57829 to MJD 58102 and calculated the time lags. 

\begin{figure*}
     \includegraphics[scale=0.6]{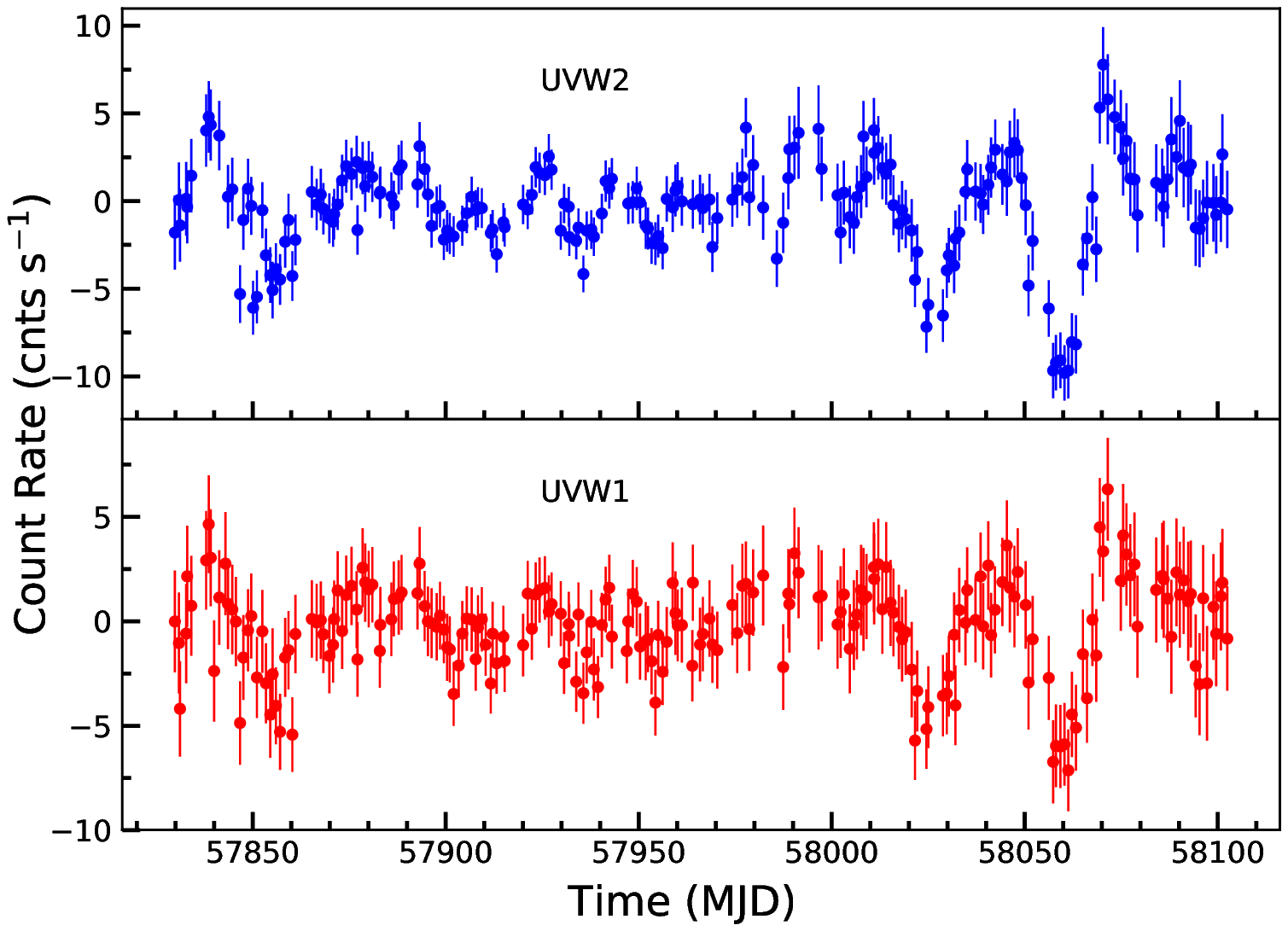}
     \includegraphics[scale=0.45]{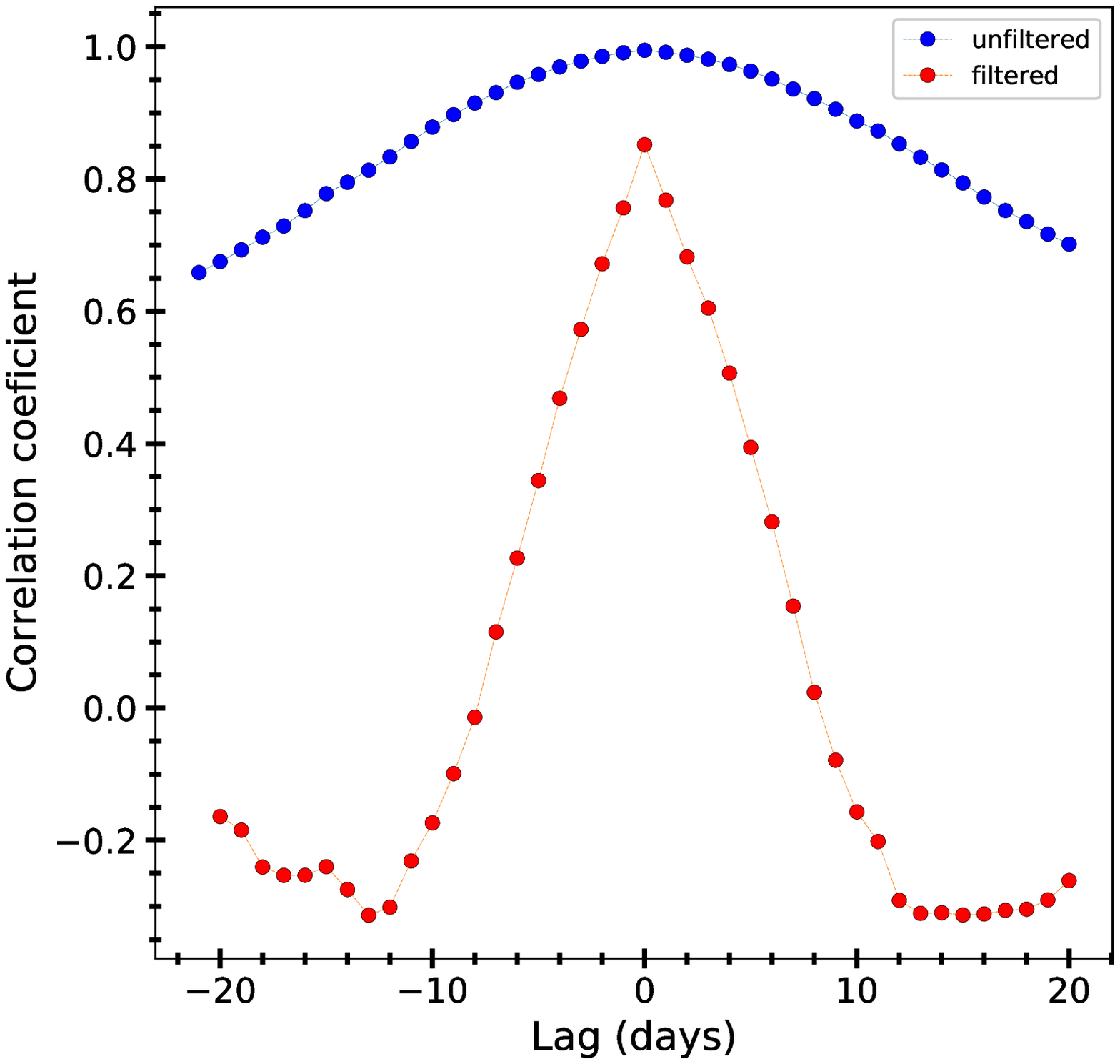}
     \caption{The light curves, filtered for the variations longer than 20 days, in UVW2 and UVW1 bands are shown in the top and bottom left panels, respectively. These filtered light curves are cross-correlated using ICCF method and resulting CCF distribution is shown in the right panel (in red circle) along with unfiltered CCF distribution (in blue circle).}
     \label{filtering}
 \end{figure*}

 We filtered out the slow variations of the order of more than 10, 15, 18, 20  and 25 days from all light curves shown in Figure~\ref{lightcur}. Although, the average cadence of the observations was about 1 day, filtering for less than 10 days was not feasible due to the gap of 6 days between two observations. Filtered light curves for the UVW2 and UVW1 bands, after removing slow variations of the order of 20 days, are shown in Figure~\ref{filtering} (left panels). 

For determining the short time delay between two filtered light curves, we applied ICCF method (described in detail in Section~\ref{subsubsec:timelag}). The estimated time delays between UVW2 and other bands after filtering have been listed in Table~\ref{lags}. The CCF lag-distribution for 20 days filtering and without filtering have been plotted in the right panel of Figure~\ref{filtering} to compare the effect of filtering on CCF distribution. It can be seen from the figure that after filtering, the distribution is significantly narrower than the unfiltered one. This indicates that the slower variations were causing the wider CCF which became narrower after filtering the long term variations from the light curves. The derived lags of X-ray emission with respect to the UVW2 emission from the unfiltered and filtered light curves clearly show the effect of the removal of slow variations (See Table~\ref{lags}). Further, to check the effect of different filtering among UV and optical bands, we filtered slow variations for different days from all the light curves and estimated time lags with respect to the UVW2 band. 

\subsubsection{Time lag estimation}
\label{subsubsec:timelag}
After removing the slow variations from the observed light curves, we measured the lags in different bands with respect to the UVW2 band. For the estimation of time lags, we used two methods that are interpolation cross-correlation function (ICCF; \citealt{1987ApJS...65....1G}, \citealt{1998PASP..110..660P}) and JAVELIN \citep{Zu_2011}. The lags estimated using these two methods were similar except that the errors calculated from JAVELIN were smaller, as previously noted by \citealp{2016ApJ...821...56F}. In the ICCF method, we assume that two light curves, say $a(t)$ and $b(t)$ are displaced by a time lag of $\tau$. First, each observed data point of the curve $b(t)$ is correlated with a linearly interpolated value of $a(t)$ at a time difference of $\tau$. Similarly, each observed data point in curve $a(t)$ is correlated with an interpolated value in $b(t)$. In this way, two correlation coefficients are calculated for each value of time lag $\tau$ and then averaged out to get the final value of the correlation coefficient. Uncertainties on the lag measurements are estimated using flux randomisation (FR) and random subset selection (RSS) methods. In the RSS method, the $N$ selections are drawn randomly from the light curve of $N$ data points for each Monte Carlo realisation, similar to the bootstraping method. The data point which is selected $M$ times, the uncertainty is decreased by $M^{1/2}$. To assess the flux uncertainty, the random Gaussian deviates are added to the data points (FR). After multiple Monte Carlo realisations, a cross-correlation function is obtained having correlation coefficient at the peak and centroid lags (usually measured above the $80\%$ of peak value). The peak and centroid lag values differ for asymmetric ICCF distribution. We performed 2000 Monte Carlo simulations for each pair of light curves of UVW2 and other bands and rejected those cross-correlation coefficients that were below 0.2. Time lags calculated with and without filtering have been tabulated in Table~\ref{lags}. \citet{Edelson_2019} calculated time lags using average count rates in the light curves for this source without removing any slow variations which are found to be consistent with the values reported in the present work (within error). However, there is a difference in the centroid value of lags in UVM2 and UVW1.               

We also applied JAVELIN method to calculate the inter-band lags. Instead of linearly interpolating between the gaps, JAVELIN models the light curves by making an assumption that the driving light curve is modeled by a damped random walk (DRW) that has been generally applied in quasars \citep{Zu_2013} and the derived light curves are related to the driving light curve via a top-hat transfer function. This method was first applied by \citet{Kelly_2009}. The time lags for X-rays, UVM2, UVW1, U, B, V bands relative to UVW2 band, without filtering are $-2.086\pm3.041$, $-0.311\pm0.218$, $0.132\pm0.196$, $1.656\pm0.337$, $1.298\pm0.404$ and $2.458\pm0.607$, respectively. After 20 days filtering, these lags were found to be $-1.232\pm0.509$, $-0.123\pm0.239$, $-0.023\pm0.266$, $0.729\pm0.488$, $1.086\pm0.639$, and $2.474\pm1.246$, respectively. These lags are consistent with that found with the ICCF method. 

 \begin{table*}
 	\caption{Time lags (in days) in different energy band relative to the UVW2 band estimated from the ICCF correlation method.}
 	\begin{tabular}{lccccccc}
 		\hline\hline
 		Band  & Wavelength (\AA)& \multicolumn{6}{c}{Time lag $\tau$ (in days)} \\
 		       &  & without filtering & 10 days & 15 days & 18 days & 20 days & 25 days\\
 		\hline\hline
         0.3-8 keV & 20.7& $3.384^{+0.976}_{-0.959}$ &$0.000^{+1.463}_{-1.000} $ &  $-0.476^{+0.971}_{-0.959} $ & $-0.476^{+0.541}_{-0.903} $ & $-0.459^{+0.538}_{-0.914}$ & $-0.068^{+0.874}_{-0.539}$  \\
 		\hline
 		0.3-2 keV & 23.8 & $2.859^{+0.944}_{-0.921}$ &$-0.013^{+1.484}_{-1.013} $ &  $-0.502^{+0.974}_{-0.949} $ & $-0.502^{+0.557}_{-0.545} $ & $-0.488^{+0.526}_{-0.541}$ & $-0.459^{+0.519}_{-0.514}$  \\
 		\hline
 		2-8 keV & 3.72 & $5.021^{+1.557}_{-2.224}$ &$0.048^{+1.549}_{-2.409} $ &  $0.022^{+1.138}_{-1.436} $ & $0.013^{+1.045}_{-1.447} $ & $0.062^{+1.065}_{-1.415}$ & $0.460^{+1.387}_{-1.025}$  \\
 		\hline
 		UVW2  &1928 &-& $-$ & $-$ & $-$ & $-$ & $-$ \\
 		\hline
 		UVM2 & 2246 &$-0.488^{+0.505}_{-0.884}$ & $0.000^{+0.516}_{-0.483}$ & $-0.041^{+0.890}_{-0.511}$  & $-0.066^{+0.879}_{-0.522}$ & $-0.447^{+0.506}_{-0.893}$ & $-0.466^{+0.497}_{-0.891}$ \\
 		\hline
 		UVW1 & 2600 & $0.481^{+0.515}_{-0.506}$ &$0.000^{+0.547}_{-0.963}$ &$0.015^{+0.526}_{-0.940}$  & $0.017^{+0.527}_{-0.929}$ & $0.015^{+0.522}_{-0.907}$ & $0.015^{+0.522}_{-0.907}$ \\
 		\hline
 		U   & 3467  &$2.017^{+0.521}_{-0.901}$ &$0.000^{+0.954}_{-2.537}$ & $0.514^{+0.970}_{-1.424}$ & $0.957^{+0.964}_{-0.979}$ & $0.982^{+0.952}_{-0.962}$ & $1.039^{+0.554}_{-0.917}$ \\
 		\hline
 		B   & 4392  &$1.452^{+0.910}_{-0.718}$ &$1.469^{+1.522}_{-5.134}$  &$1.504^{+1.493}_{-2.036}$  & $1.449^{+1.446}_{-1.501}$ & $1.495^{+1.090}_{-1.032}$ & $1.485^{+0.991}_{-0.997}$\\
 		\hline
 		V   & 5468  &$1.622^{+0.980}_{-0.922}$ & $1.456^{+2.456}_{-3.562}$ & $2.474^{+2.474}_{-2.507}$ & $2.473^{+2.472}_{-2.062}$ & $2.523^{+2.520}_{-2.411}$ &  $2.485^{+1.971}_{-1.960}$\\
 		\hline\hline
 		\multicolumn{8}{c}{Soft X-ray (0.3-2 keV) lag wrt Hard X-rays (2-8 keV)}\\
 		\hline\hline
 		  &    & $-0.487^{+0.503}_{-0.492}$ & $0.000^{+0.000}_{-0.000}$ &  $0.000^{+0.462}_{-0.452}$ & $0.000^{+0.466}_{-0.453}$ & $0.000^{+0.469}_{-0.454}$ & $-0.003^{+0.468}_{-0.455}$\\
 		\hline
 	\end{tabular}
 	\label{lags}
 \end{table*}

We derived the lag spectrum, i.e. lag as a function of wavelength for unfiltered and filtered (for 18 days filtering) light curves and showed it in the left and right panels of Figure~\ref{lagplot}, respectively. We then used a phenomenological power law model to model the lag spectrum. An expression of such power law model in terms of lags, i.e. the time-delay between a reference wavelength ($\lambda_0$) and other wavelength ($\lambda$) can be given as $$\tau=\alpha \left[ \left( \frac{\lambda}{\lambda_0} \right)^{\beta} -1 \right]$$ where $\tau$, $\alpha$ and $\beta$ are time delay between reprocessing and reprocessed emission, power-law normalization and power-law index, respectively. In the standard accretion disk theory, the value of power-law index $\beta$ is considered to be 4/3 \citep{1999MNRAS.302L..24C, 2007MNRAS.380..669C}. We modelled the lag spectrum with the power-law model in the following ways: (a) varying normalization $\alpha$ and power-law index $\beta$ with and without hard and soft X-ray lags, and (b) by fixing the value of $\beta$ as $4/3$ and varying normalization only. Since there was an excess lag in U band for unfiltered case, it was removed from the lag spectrum. Here, the UVW2 band is considered as the reference waveband. The parameters obtained from fitting the lag spectrum for unfiltered light curves are (a) $\alpha$ = 0.14$\pm$0.75, $\beta$ = 2.71$\pm$5.08 (including all the wavebands), and $\alpha$ = 1.49$\pm$5.71 $\beta$ = 0.78$\pm$2.11 (excluding X-ray and U bands), and (b) $\alpha$ = 0.51$\pm$0.44. After 18 days filtering of light curves, the parameters are (a) $\alpha$ = 0.26$\pm$0.15, $\beta$ = 2.28$\pm$0.59 (including all wavebands), $\alpha$ = 0.20$\pm$0.19, $\beta$ = 2.52$\pm$0.98 (excluding X-ray bands), and (b) $\alpha$ = 0.61$\pm$0.12. The lag spectrum along with fitted power law models (as described above) without filtering and 18 days filtering of light curves are shown in the left and right panels of Figure~\ref{lagplot}, respectively. From the figure, it is clear that there is a large discrepancy between the model and observed lag even after filtering the light curves for long term variations.

\section{Discussion}
\subsection{Correlations observed in the X-ray and UV/optical emission}
\label{subsec:correlation}
From the analysis, we found that the average spectrum over a duration of thirteen years shows the presence of a complicated soft X-ray excess over the power-law continuum. The soft X-ray excess is described well with a combination of two thermal components with temperatures of $kT_{\rm bb1}\sim 120$ eV and $kT_{\rm bb2}\sim 460$ eV. The low temperature thermal component is normally required in Seyfert~1 AGNs \citep{1999ApJS..125..317L}. However, the warm thermal component may be an analogy to the warm Comptonization, as described by \citet{2012MNRAS.420.1848D}, to explain the soft X-ray excess. According to this model, the warm Comptonizing plasma ($\sim 0.2-0.5$ keV) is embedded inside the optically thin and hot Comptonizing plasma ($\sim 100$ keV). The soft X-ray excess due to warm Comptonization was also reported in this AGN by \citet{2011A&A...534A..39M}. Such a geometry of disk/X-ray plasma may be supported due to the observed correlation between the soft X-ray and hard X-ray emission ($\rho\sim0.82$, log$p \sim-71$). In that case, the variability amplitude observed in the UV and soft X-ray bands should be, at least, equal to that of the highly variable power-law continuum flux. From the estimation of fractional variability, we found a low variability amplitude of the power-law continuum flux ($\sim 15\%$). This suggests that the hot X-ray plasma may not be lying over the warm Comptonizing plasma, but situated at a different location closer to the SMBH. Further, the correlation between the derived spectral components such as power-law flux and BB flux and BB temperature and BB flux are weak or moderate while other combinations do not show any correlation (see Figure~\ref{par_corr}). The observed weak anti-correlation between the power-law photon index and blackbody temperature, though both the components overlap only in a narrow energy range (below 2 keV), is apparently due to the degeneracy between both the parameters. Therefore, the emitting regions are likely to be partly interacting or distinct. At the same time, the variability observed in the UV emission, which is supposed to be radiated from regions close to the SMBH as compared to the optical emission, is also higher than that observed in the hard X-ray emission. However, the hard X-ray emission variability is higher relative to the changes found in the V-band. This also suggests that the UV and hard X-ray emission may be associated with disjoint regions or partly interacting regions. Similarly, the C3PO analysis gives a negative offset for the soft X-ray and UV/optical bands (except the V band) while taking the hard X-ray band as the abscissa. This indicates that these bands are more variable than the hard X-ray.

Weak/moderate correlations between the X-rays and the UV/optical emission may be interpreted as due to the presence of fluctuations of different time scales. These fluctuations may be associated with the changes in the accretion rates according to the fluctuation propagation model proposed by \citet{1997MNRAS.292..679L}. Typical time scale of variations expected for the UV/optical bands due to changes in the accretion rates are of the order of thousands or more years for AGNs like Mrk~509. These time scales are estimated for 1H~0419-577 by \citet{2018MNRAS.473.3584P}, which has a similar supermassive black hole mass ($\sim~10^8~\rm M_{\odot}$).       
 
The soft excess in AGNs is often explained by the relativistic blurred disk reflection model \citep{2000PASP..112.1145F, 2009Natur.459..540F}. According to this model, the accretion disk is illuminated by the X-ray power-law continuum emission from the corona, producing a reflected spectrum. Reflection of continuum photons from the relativistically rotating inner disk gives rise to several smeared emission lines which look like a hump (an excess over the continuum emission) below $\sim$ 2 keV. In the case of Mrk 509, however, the soft X-ray leads the hard X-ray (power-law) emission as shown in Table~\ref{lags}, which is not compatible with the reflection model. Secondly, the low variability in hard X-ray and weak/moderate correlation between hard X-ray and UV/optical bands make it unlikely that the UV/optical emissions are driven by reflection.

\subsection{Possibility of the X-ray reprocessing}
\label{subsec:reprocessing}
\begin{figure*}
    \includegraphics[scale=0.547]{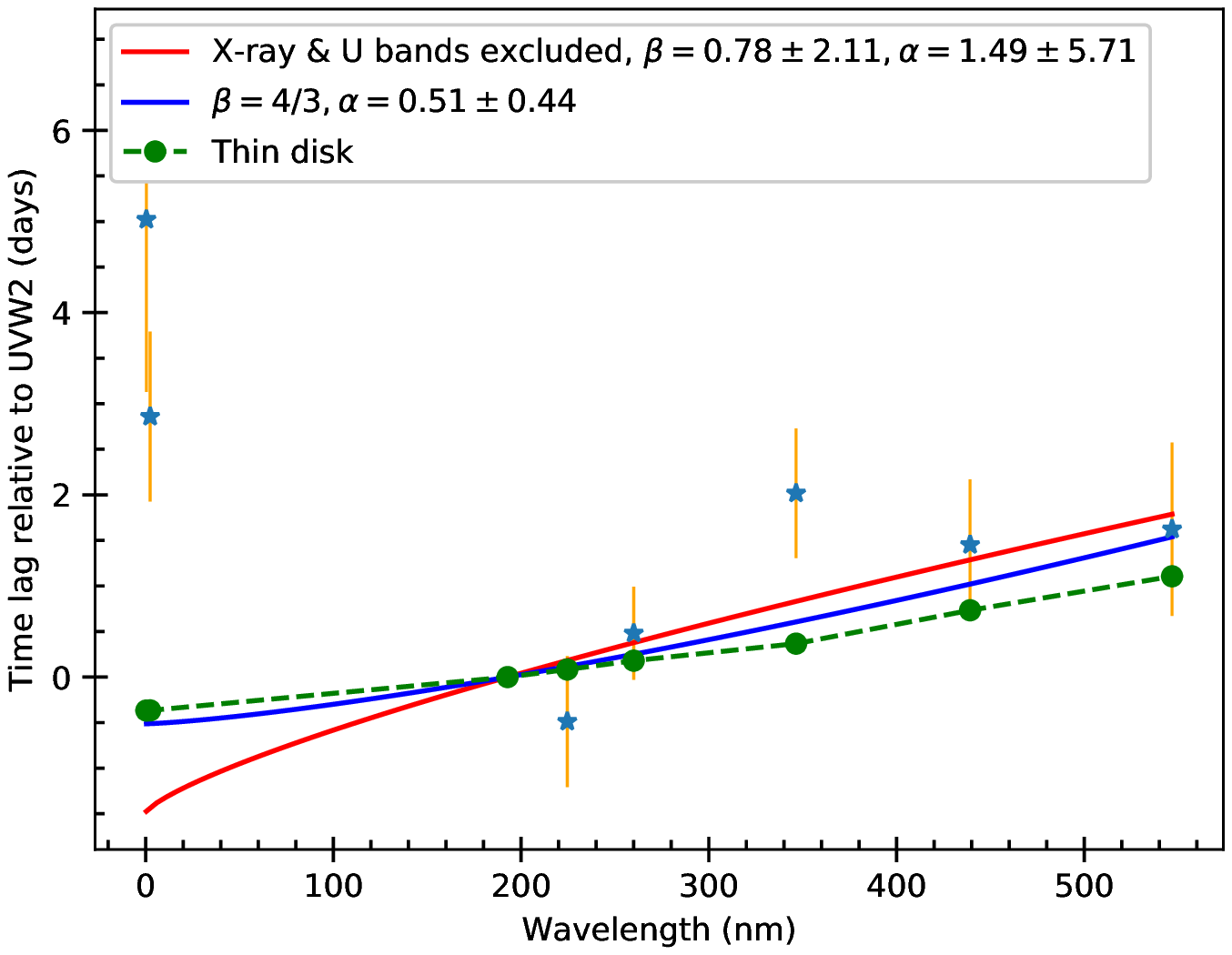}
     \includegraphics[scale=0.547]{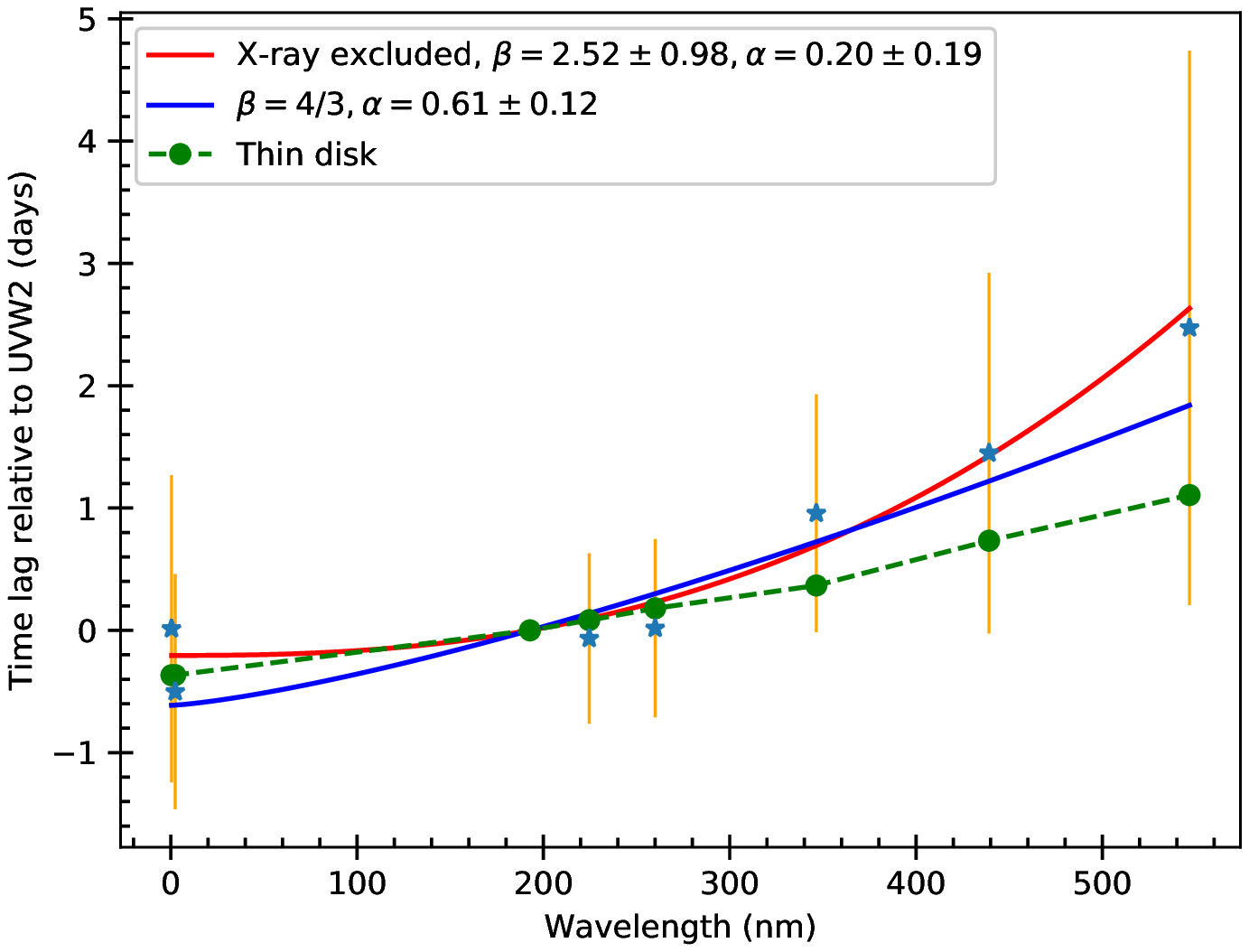}
     \caption{Wavelength dependent lag spectrum modeling is shown for without filtering (left panel) and 18 days filtering (right panel) of light curves (i) with power law model $\tau\propto \lambda^{\beta}$ excluding X-ray data points (and U in the case of without filtering) and extrapolated down to X-ray (red solid line) (ii) with simple 4/3 power law (blue solid line) (iii) Theoretically estimated time-delay with respect to UVW2 band (green dashed line). Blue stars are the time lag values calculated from ICCF method.}
     \label{lagplot}
 \end{figure*}

X-ray reprocessing phenomenon is a process in which X-ray emission from the optically thin and hot plasma is reprocessed in the distant accretion disk to re-emit in the longer wavelengths \citep{1991ApJ...371..541K, 1999MNRAS.302L..24C}. Confirmation of the X-ray reprocessing was first observed from UV/optical to infrared bands in a Seyfert~1 AGN NGC~2617 by \citet{Shappee_2014} and the lags at the longer wavelengths were found to be consistent with the predictions of the X-ray reprocessing in the standard accretion disk. In recent times, NGC~5548 has been the most studied AGN for the reverberation mapping of the accretion disk \citep{2015ApJ...806..129E, 2016ApJ...821...56F}, and most of the studies hint towards a larger size of the accretion disk. Theoretically, in a geometrically thin and optically thick accretion disk, which is heated internally by viscous dissipation and externally by irradiation of the central UV/X-ray source, the temperature at a radius $R$ from the center is given by 
 \begin{equation}
 	T(R) = \Bigg( \frac{3GM \dot{M}}{8\pi \sigma R^3} + \frac{(1-A) L_x H}{4\pi \sigma R^3} \Bigg)^{1/4} 
 	\label{temp}
 \end{equation}

where $G$, $M$, $\dot{M}$, $L_x$, $A$, $H$ and $\sigma$ are the gravitational constant, mass of the central SMBH, mass accretion rate, luminosity of the heating radiation, albedo of the accretion disk, height of the central heating source and Stefan-Boltzmann constant, respectively \citep{2007MNRAS.380..669C}. In this equation, the effects of inclination angle, relativistic effects in the inner part of the disk and the inner edge of the disk have been neglected. Following the steps given in \citet{2016ApJ...821...56F}, the time delay $\tau$ relative to the reference time-delay $\tau_0$ corresponding to reference wavelength $\lambda_0$ is given by 
 \begin{equation}
 	\tau-\tau_0 = \frac{1}{c} \Bigg(X\frac{k\lambda_0}{hc} \Bigg)^{4/3} \Bigg( \frac{3GM\dot{M}}{8\pi\sigma} + \frac{(1-A) L_x H}{4\pi \sigma} \Bigg)^{1/3}\times \Bigg[\Bigg(\frac{\lambda}{\lambda_0}\Bigg)^{4/3}-1 \Bigg]
 	\label{timedelay}
 \end{equation}

where, $X$ is the multiplicative factor that takes into account systematic issues in the conversion of temperature $T$ to wavelength $\lambda$ for a given $R$. The value of this factor is 4.87 when the temperature corresponding to the observed emission wavelength is given by Wein's law and 2.49 when flux-weighted radius is used. In the flux-weighted case, the temperature profile of the accretion disk is described by the Shakura \& Sunyaeav disk ($T \propto R^{-3/4}$). In both the cases, the disk is assumed to have a fixed aspect ratio all over. Eqn.~\ref{timedelay} can be simplified by taking $(1-A)L_xHR = \kappa GM\dot{M}/2R$, where $\kappa$ is the local ratio of external to internal heating and independent of radius. Using luminosity of Mrk~509 $L_x$ as 10$^{44}$ erg s$^{-1}$ \citep{10.1111/j.1365-2966.2008.14108.x}, coronal height $H$ as 1.53 $r_g(=GM/c^2)$ \citep{Garc_a_2019} and assuming albedo $A$ equal to 0.2 (as used in several AGNs e.g. \citet{10.1093/mnras/stu1636}, \citet{10.1093/mnras/stx3103}), the value of $\kappa$ comes out as $\sim$0.016. The smaller value of $\kappa$ indicates that the contribution towards time delay due to external heating is insignificant (0.5\%, estimated from Eqn.~\ref{timedelay}) compared to viscous heating. Therefore, using  Eqn.~3 from \citet{Edelson:2017jls}, the time delays with respect to UVW2 ($\tau_0=0$) band can be expressed as 
 
 \begin{equation}
 	\tau = 0.09 \Bigg(X \frac{\lambda}{1928\angstrom} \Bigg)^{4/3} M_8^{2/3} \Bigg(\frac{\dot{m}_{edd}}{0.10} \Bigg)^{1/3}\times \Bigg[\Bigg(\frac{\lambda}{\lambda_0}\Bigg)^{4/3}-1 \Bigg] days
 \end{equation}

where, $M_8$ and $\dot{m}_{edd}$ are mass of the black hole in units of $10^8~M_{\odot}$ and the Eddington ratio ($\dot{M}$/$\dot{M}_{edd}$). This equation assumes negligible external heating by the UV/X-ray source compared to the internal viscous heating and radiation efficiency to be 0.1. We use the mass of the central SMBH of Mrk~509 to be $1.43\times10^8 ~M_\odot$ \citep{Peterson_2004}, $\dot{m}_{edd}$ equal to 0.0951 \citep{10.1111/j.1365-2966.2008.14108.x} and $X$=2.49 for flux-weighted radius. Using the above values, the time-delays with respect to the UVW2 band are estimated to be $-0.367, -0.366, 0.083, 0.180, 0.367, 0.733$ and $1.106$ days for hard X-rays, soft X-rays, UVM2, UVW1, U, B and V bands, respectively. In our analysis of unfiltered light curves of Mrk~509, we found delayed emission in hard X-ray, soft X-ray, UVW1, U, B and V bands with respect to the emission in UVW2 band (Figure~\ref{lagplot} \& Table~\ref{lags}) while the delayed emission in the X-ray reprocessing is expected to be in the longer wavelength bands compared to the X-ray emission. Therefore, the observed positive lags in the X-ray bands are unexpected while considering the reprocessing scenario. 

The estimated lags from the observed light curves may be affected due to the slow variations, for example, due to the fluctuations in the accretion flow. The reprocessed variable component in longer wavelength is diluted with the intrinsic emission due to the viscous heating in the accretion disk. Therefore, a possible origin of the variability in longer wavelengths could be a combination of two effects: reprocessing and propagating fluctuations. In the propagating fluctuation scenario, one expects the longer wavelength emission leading to the shorter wavelength emission. As the fluctuations propagate inwards through the accretion flow, it modulates the UV/optical and X-ray emission. Since the X-ray emitting region is very close to the central engine, it shows rapid fluctuations. These X-rays are again reprocessed to give longer wavelength emission. Therefore, there is always a tussle between the reprocessing and propagating fluctuations and the resultant time lag depends on the dominant process. Therefore, the geometry of the corona and the accretion disk are complex and both are not so simple as assumed in the lamp-post model. After removing the slow variations of about 18 days, the trend in the estimated lags appear to be the same as the delayed hard X-ray emission as well as the optical emission (right panel of Figure~\ref{lagplot}).

\subsection{Excess lags in U and X-ray band with respect to UVW2 band}  
\label{subsec:excesslag}
The lags derived without removing the slow variations are shown in the left panel of Figure~\ref{lagplot}. This shows clear "excess" lags in U as well as in X-ray bands with respect to the UVW2 band when fitted with the power law model $\tau \propto \lambda^{4/3}$. Excess lag in U band was first reported by \citet{2001ApJ...553..695K} in NGC 5548 and then in later studies \citep{2015ApJ...806..129E, 2016ApJ...821...56F, 10.1093/mnras/stx3103, 2019BSRSL..88..143N} in different AGNs. \citet{Cackett_2018} also found a strong excess in the Balmer jump region in NGC~4593. \citet{2001ApJ...553..695K} predicted that there is a significant contribution of diffused emission from the BLR in UV bands. The wavelength-dependent lag across UV-optical energy range can imitate the lag spectrum originated as a result of X-ray reprocessing from the accretion disk. Similar results were found by \citet{10.1093/mnras/sty2242} and they emphasized on the importance of diffuse continuum contribution while inferring the disk sizes on the basis of inter-band continuum delays. This feature has been observed in many Seyfert~1 galaxies, as shown by \citet{Edelson_2019} in four AGNs, including Mrk~509. \citet{10.1093/mnras/staa2365} modelled the lag spectrum of Fairall~9 by taking into account the contribution from the BLR (Figure~6 of their paper). Their modelling was based on the previous work on NGC~5548 by \citet{10.1093/mnras/stz2330}. On the basis of these results, it can be safely inferred that the BLR contribution in the inter-band lag spectra is unavoidable. As a result, the variability within the UV/optical can be inferred as the mixing of two effects: emission from standard Shakura-Sunyaev disk and diffused continuum emission from the BLR as discussed above.

In the simple picture of reprocessing model, X-ray emitting central corona illuminates and heats the accretion disk. The reprocessing model predicts that when $\tau \propto \lambda^{4/3}$ relation in UV/optical is extrapolated down to X-rays, it should fit the data well. But Figure~\ref{lagplot}, generated from unfiltered light curves for long term variations, shows a huge discrepancy. Even after filtering, although excess lag was decreased significantly, the lags in hard X-rays still appear to be higher, which can not be explained by a simple reprocessing model.

\subsection{Complex nature of the accretion disk}
\label{subsec:complexnature}
As the X-ray reprocessing phenomenon is unable to explain the observed lag spectrum successfully, this compels us to look for alternative models in explaining the observed correlated variability in the X-ray/UV/optical wave-bands. One such model is given by \citet{10.1093/mnras/stx946} in which the excess lag of X-ray relative to UV band in NGC~5548 is explained. According to this model, the inner part of the accretion disk is called the Comptonization region, which is "puffed up" and emits in EUV energy band. This region is heated by the hard X-rays coming from the central corona, producing heating waves which dissipate outwards. In response to these heating waves, the outer region of the disk expands and contracts and produces continuum lags. The model suggests the shorter wavelength lead the longer wavelength variations, which is not the case with Mrk~509 even after filtering. So this model does not explain the short term variability seen in Mrk~509 for this particular duration of observations. Using \xmm{} and \swift{} observations of Mrk~509 for a few days, \citet{2011A&A...534A..39M} inferred that the soft X-ray excess below 2 keV is produced by the Comptonization of the seed photons from the UV-optical disk into a warm and optically thick corona. This result seems to be consistent with the positive lag between UVW2 and X-rays found in our analysis. 

The best-fitted value of power-law normalization $\alpha$ gives the size of the emitting region of reference wavelength, i.e. 1928\AA (the central wavelength of the UVW2 band) if we assume the face-on standard disk. The size of the emitting region for the UVW2 emission can be estimated to be $0.61\times24\times60\times60\times 3\times10^{10}=1.58\times10^{15}$cm which is equivalent to $\sim$80$R_{g}$. Therefore, the size for the UVM2 and UVW1 emission would be similar as these bands show the time delays consistent with zero for the UVM2 and UVW1 bands with respect to the UVW2. In the case of variability, the fractional variability found in the UVW2 band is marginally higher than that seen in the UVM2 and UVW1 bands. The lags consistent with zero and similar amplitude of the variability in the UV bands suggest that the disk does not appear to have a smooth surface. Though, the $\beta=$4/3 wavelength rule does not fit the lag spectrum well when the X-ray and U band lags are included; the spectrum is well described by this rule on excluding these excess lags. This implies that the reprocessed optical emission is likely caused by the UV emission with some delay.  
Excluding the X-ray lags, the fits corresponding to $\beta=2.52\pm0.98$ and the $\beta=4/3$ values describe the lag spectrum well (right panel of Figure~\ref{lagplot}). These fits suggest that the UV emission is probably reprocessed in optical as well as in X-ray bands. 
  
The fitted $\beta$ values ($0.78\pm2.11$ without X-ray and U; $2.52\pm0.98$ after filtering) infer a complex structure of the accretion disk and even rules out a shell-like geometry ($\tau\propto\lambda^2$) of emitting/reprocessing regions around the SMBH. Therefore, the actual disk is likely to be partly puffy, warped and inhomogeneous in nature and thus, the standard accretion theory seems not viable in the case of Mrk~509. The reason for getting lower lags after the process of filtering slow variations can not be explained simply. In this process of removing of the slow components, it is not clear about the increased lags of V band, i.e. $\sim 1$ day for 10 days filtering and $\sim 3$ days for 20 days filtering. To answer the above questions and exploring such a complex nature of the accretion disk of Mrk~509 require a longer monitoring program with a good cadence of observations compared to the available one.
\section{Summary} 
\label{sec:summary}
We analyzed the X-ray and UV/optical data of \swift{} observatory spanning 2006 to 2019 to study the spectral and timing variability properties of Mrk~509. We applied different techniques to find out the correlation and time lags between different energy bands. We could see the correlation among all the energy bands by visual inspection. To quantify the correlation, we used count-count (or flux-flux) correlation with positive offset method. The time lags between various energy bands were estimated using ICCF and JAVELIN methods after filtering long variability components using the LOWESS method. The main results of our analysis are as follows:

 \begin{enumerate}
 	\item Average spectrum over the entire duration shows that the soft X-ray excess is well described by two blackbody components with temperatures $kT_{\rm bb1}\sim120$ eV and $kT_{\rm bb2}\sim460$ eV. 
	\item We found a moderate/weak correlation between (i) the power-law flux and BB flux, and (ii) the blackbody temperature and BB flux. 
 	\item Mrk~509 shows strong variability on short as well as long term time scale (see variability amplitudes in Section~\ref{subsubsec: varaibility}) in all bands. Interestingly, the X-ray power-law continuum shows smaller variability amplitude ($\sim$15\%) compared to the soft X-ray and UV bands ($\sim$18-28\%).  
 	\item The variability process in this AGN seems to be non-stationary (Figure~\ref{stationarity}). However, no spectral transition is observed over the years, as can be seen in Figure~\ref{KT-index} where the power-law photon index and blackbody temperature do not show any significant change.
 	\item The value of Pearson's correlation coefficients for the UV/optical bands ($\rho = 0.95-0.99$ with respect to the UVW2 bands and $\rho = 0.40-0.53$ with respect to the hard X-ray) and corresponding $p$ values show that UV and optical bands are significantly correlated as compared to that of the X-ray emission and UV/optical bands. The hard X-ray emission is found to be well correlated with the soft X-ray emission ($\rho$=0.82) while the hard X-ray and soft X-ray emission are correlated to the UVW2 emission with $\rho$ values of 0.51 and 0.75, respectively.
	\item The positive offset resulting from the linear fit in the CP3O method with respect to UVW2 band suggests that the emission in UV and optical bands is less variable, while the negative offsets in the soft bands (soft X-ray/UV/optical) with respect to the hard X-ray power-law continuum flux infer relatively more variabilities in the soft bands. 
 	\item The time lags were estimated with respect to UVW2 energy band using ICCF and JAVELIN methods. The lags estimated from these two methods were comparable (within error). Before filtering the light curves for long-term variabilities, there was an excess lag in X-ray band and U-band. After filtering for 18 days, the lag got reduced significantly. 
 	\item The observed lags were found to increase with wavelength and were fitted with power law model by (i) keeping power-law parameters (i.e., $\alpha$ \& $\beta$) free  with and without excluding hard and soft X-ray (ii) fixing $\beta$ at $4/3$ according to Shakura-Sunyaev accretion disk geometry.
 	\item Our results favour the warm Comptonization region, located above the inner part of the accretion disk as the origin of the observed soft X-ray emission from Mrk~509. However, zero time lag and marginal variability difference between UV bands suggest that the UV emitting region in the disk is not smooth, instead a stratified region.
\end{enumerate}

\begin{acknowledgements}
We thank the reviewer for his/her useful comments/suggestions which improved the paper significantly. The research work at PRL is funded by the Department of Space, Government of India. MP thanks the UGC, India for financial support through the DSKPDF fellowship grant no. BSR/2017-2018/PH/0111. For this research work, the software and online tools provided by the High Energy Astrophysics Science Archive Research Center (HEASARC) online service, maintained by the NASA/GFSC and the High Energy Astrophysics Division of the Smithsonian Astrophysical Observatory have been used. This work made use of data supplied by the UK Swift Science Data Centre at the University of Leicester.
\end{acknowledgements}

\bibliographystyle{pasa-mnras}
\bibliography{ref}

\end{document}